\documentclass[a4paper,11pt]{article}
\usepackage{jcappub} 
\usepackage{lineno}
\usepackage{graphicx}
\usepackage{subcaption}

\title{The Dusty Selection Function: How Reddening-Map Errors Bias Stage-IV Cosmology}

\author[a,b]{T. Karim\note{Corresponding author.}}
\affiliation[a]{Dunlap Institute for Astronomy \& Astrophysics, University of Toronto, 50 St. George St, Toronto, ON M5S 3H4, Canada}
\affiliation[b]{Center for Astronomy, Space Science and Astrophysics, Independent University, Bangladesh, Dhaka 1229, Bangladesh}

\author[c]{A. S. Maniyar}
\affiliation[c]{Laboratoire d'Annecy de Physique Th\'eorique (LAPTh), CNRS/USMB, 9 Chemin de Bellevue BP110, Annecy, F-74941, ANNECY CEDEX, France}

\emailAdd{tanveer.karim@utoronto.ca}

\abstract{
A limiting factor for the cosmological reach of Stage-IV galaxy surveys is how well we know the galaxy selection function. Galaxy samples are selected on photometry that has been corrected for Milky Way extinction, so any error of the reddening map is imprinted on the density of the sample. We measure this imprint in the photometric emission-line galaxies of the DESI Legacy Imaging Surveys, which are selected with the Schlegel, Finkbeiner \& Davis (SFD98) map. An error of 10 mmag in $E(B-V)$ changes the density of this sample by 9.5\%, and that of the DESI DR1 spectroscopic sample by about 11\%. We then measure how much of this imprint survives the corrections in use today. In the photometric sample, a correction with an absolute reddening map leaves 60--80\% of its amplitude. In the DESI DR1 spectroscopic emission-line galaxy sample, the imaging weights estimated with reddening-difference templates still leave 30--50\% of the amplitude. Two far-infrared maps---SFD98 and \textit{Planck} GNILC---also contain large-scale structure, which we detect through their correlation with CMB lensing. For SFD98 this adds at least 5--6\% to the galaxy--lensing cross-spectrum. We then study what this means for a Rubin LSST $\times$ Simons Observatory analysis. For a sample with the same response, $f_{\rm NL}$ and $\Omega_m$ are biased by 1--2$\sigma$ after a correction with an absolute reddening map. The bias of $\sigma_8$ exceeds 1$\sigma$ once the photometric redshift scatter is known to 10\% of its value, and reaches 2--3$\sigma$ when it is known exactly. The neutrino mass is biased by 1--2$\sigma$ in either direction, including to a formally negative value. A bias below 0.5$\sigma$ in $f_{\rm NL}$ requires a response about half that of these galaxies. We recommend that the response to a reddening error be measured for every tomographic bin, and that samples be selected with a reddening map that has been tested for extragalactic signal.}

\begin{document}
\maketitle
\flushbottom

\section{Introduction}
\label{sec:intro}

Wide-field surveys such as the Dark Energy Spectroscopic Instrument (DESI) \cite{desi-dr1-data-2026}, the Legacy Survey of Space and Time (LSST) at the Vera C. Rubin Observatory \cite{lsst-main-ivezic}, the Simons Observatory (SO) \cite{so-forecast-2019}, the Subaru Prime Focus
Spectrograph \cite{pfs-overview-2014}, SPHEREx \cite{dore-spherex-2014}, \textit{Euclid} \cite{euclid-overview-2025} and \textit{Roman} \cite{roman-overview-2019} are opening the way to test cosmological models to high precision. The clustering of galaxies and its cross-correlation with CMB lensing measure the amplitude of matter fluctuations, $\sigma_8$, the matter density, $\Omega_m$, the sum of the neutrino masses, $\Sigma m_\nu$, and the amplitude of primordial non-Gaussianity, $f_{\rm NL}$ \cite{so-forecast-2019, lsst-desc-srd}. These parameters draw on different parts of the measurement. The amplitude of the spectra constrains $\sigma_8$ and $\Sigma m_\nu$, while their shape on large scales constrains $\Omega_m$ and $f_{\rm NL}$.

Among these, $f_{\rm NL}$ is one of the key science drivers of the current era, because surveys now reach the volume needed to probe the largest scales. Multi-field models of inflation predict a primordial non-Gaussianity that changes the abundance of halos and gives the galaxy bias a dependence on scale that grows as $k^{-2}$ \cite{dalal2008-fnl}. While single-field slow-roll inflation predicts $f_{\rm NL} \sim 0.01$ \cite{maldacena2003}, multi-field
models predict $f_{\rm NL}$ of order one \cite{alvarez2014-fnl-forecast, dePutter2017-fNL-forecast}, so a measurement with $\sigma(f_{\rm NL}) \sim 1$ can rule out whole classes of models. The best current constraint comes from \textit{Planck}, $f_{\rm NL} = -0.9 \pm 5.1$ \cite{planck2020-fnl-best}, and is limited by cosmic variance. Galaxy surveys can do better because they measure many more modes. The DESI analysis of luminous red galaxies and quasars gives $f_{\rm NL} = -3.6^{+9.0}_{-9.1}$ \cite{edmund2024-desi-dr1-fnl}, and cross-correlations of photometric samples with CMB lensing currently reach $\sigma(f_{\rm NL}) \approx 20$--$30$ \cite{bermejo2025-dr9-lrg-planck-cmb-lensing, krolewski2024-qso-dr9-planck, mehdi2024-lrg}.
The latter are expected to improve substantially, since the cross-correlation cancels part of the cosmic variance \cite{schmittfull-seljak-variance-cancel-2018} and since samples at $z > 2$ carry a stronger signal \cite{ferraro2019-z-over-2}. The signal, however, lives on the largest scales, and these are the scales where the observed galaxy density is hardest to trust.

For all these measurements, we are no longer limited by data volume, but rather how well we know the \textit{selection function}. Selection function encodes how the observed density of a sample varies across the sky for reasons other than cosmology. Variations in depth and seeing, stars and Milky Way dust all change the observed density, and they do so mostly on large scales, where they add power in the same way as a positive $f_{\rm NL}$ \cite{mehdi2020-decals7-img-sysnet}. The current approach to this issue, known as imaging systematics or
observational systematics, is to regress the observed density on maps of these effects, with linear models or neural networks, and to weight the galaxies accordingly \cite{mehdi2020-decals7-img-sysnet, alberto-2024-elg-sysnet}. The samples that reach the highest redshifts are often more affected by this issue. Apart from bright quasars, most high-redshift tracers are star-forming galaxies selected near the faint limit of the imaging, often in the bluest bands, e.g., the emission-line galaxies (ELGs) of DESI and the Lyman-break galaxies of Rubin LSST. Thus, their density is the most sensitive to any error in the photometry.

Milky Way dust is a particular case among these effects, because we do not observe it directly. We only have maps that trace it, and these maps do not agree with each other. The map of \cite{sfd98} (SFD98), which is built from far-infrared emission, has been the standard choice in cosmology for many years as it gives all-sky, high-resolution coverage. However, it is known to contain emission from distant galaxies \cite{chiang-menard-2019, csfd}, and its conversion from emission to reddening depends on the properties of the dust grains, which vary across the sky \cite{dennis-lee-2025-end-of-fir}. Reddening
maps built from the colors of stars are now available over part of the sky \cite{rongpu-desi-dustmap-zhou, nayantara-mudur-dust}, and the DESI BAO and full-shape analyses \cite{desi2025-sample-characteristics} use the difference between each of the two DESI stellar reddening maps \cite{rongpu-desi-dustmap-zhou} and SFD98 as templates in their regression \cite{alberto-2024-elg-sysnet}. The effect of photometric calibration errors of this kind on clustering was studied in general by \cite{huterer-cunha-fang-2013}.

In this paper we address one specific way in which the reddening map enters a galaxy sample. Dust affects a galaxy sample twice. The better-known effect is that wherever the extinction is larger, fewer faint objects are detected. To correct for the missing galaxies, statistical weights are computed using a regression between the observed number density and dust column templates. There is another effect, via the \textit{selection cuts}. A galaxy sample is defined through cuts in the color-magnitude space, and these cuts are applied to magnitudes that have already been corrected with a reddening map. When that map is wrong, the corrected magnitudes are wrong, and galaxies cross the
selection boundary in both directions. The error of the map is then imprinted on the density of the sample, with an amplitude set by how steeply the selection responds to a change in magnitude. This second effect is the subject of this paper.

We organize our study around three questions, with the photometric ELG sample of \cite{karim2025-elg-cmb-lensing} from the DESI Legacy Imaging Surveys DR9 as a test case. First, \textit{how strongly does a sample respond to an error of the reddening map}? We measure the response directly from the catalog and find that an error of $10$~mmag changes the density of the ELG sample by $9.5\%$. Second, \textit{what does the error of the map contain}? We separate a Galactic part, which is independent of the large-scale structure, from an extragalactic part, which follows the large-scale structure and therefore also enters the cross-correlation with CMB lensing. Third, \textit{how much of the imprint survives the corrections in use today}? We measure what is left after a regression on each of five reddening maps. We also critically evaluate how much of the imprint the DESI DR$1$ weights remove,
using reddening-difference templates that differ from the ones in the DESI regression.

We then propagate what is left to a Rubin LSST $\times$ SO analysis for a sample ($z \gtrsim 0.7$) with the same response. We find that the Galactic part gives a systematic bias in the parameters that are sensitive to the shape of the angular power spectrum, $f_{\rm NL}$ and $\Omega_m$. The extragalactic part biases $\sigma_8$ and the neutrino mass by an amount that depends on how well the redshift distributions are known. For $\sigma_8$, a poorly known redshift distribution absorbs the imprint, and a well-known one reveals the systematic bias. The neutrino mass can be biased in either direction, including to formally negative values. We turn these results into requirements on the response of a sample and on its redshift calibration, and into a list of steps for an analysis.

Three things make our paper timely. First, the problem is already present in current data. In the DESI photometric ELG sample, changing only the reddening map given to the regression changes the tension with \textit{Planck} in $\sigma_8$--$\Omega_m$ from $4.29\sigma$ to $3.07\sigma$, even with a conservative scale cut of $\ell_{\rm min} = 100$~\cite{karim2025-elg-cmb-lensing}. Second, SFD98 is still the default reddening map for selecting galaxy samples, as in the DESI target selection, and the next generation of photometric samples will reach fainter limits in bluer bands, where the response to a reddening error is largest. Third, the stellar reddening maps tested here do not cover the southern sky, where most of the overlap of Rubin and SO lies.

The paper is organized as follows. Section~\ref{sec:sel-func-impact} presents our model of the imprint. We define the response of a sample to a reddening error, show how a regression on a template acts on the imprint, and derive what is left when the template is imperfect. Section~\ref{sec:data} describes the galaxy samples, the reddening maps and the CMB lensing maps that we use.

Section~\ref{sec:results-reddening} contains our measurements. We measure the response of the ELG sample, test the reddening maps for extragalactic signal with CMB lensing, and measure what is left of the imprint after a regression on each reddening map and after the DESI weights. Section~\ref{sec:forecast} describes the Rubin LSST $\times$ SO forecast and how the measured terms are injected into it. Section~\ref{sec:results-forecast} gives the resulting systematic biases in $f_{\rm NL}$, $\sigma_8$, $\Omega_m$ and $\Sigma m_\nu$,
and the requirements on the response of a sample and on its redshift
calibration. Finally, we summarize our findings and give recommendations for future analyses in Section~\ref{sec:discussion-conclusion}. A reader interested mainly in the implications for Rubin and SO can start at Section~\ref{sec:forecast}, which restates the measured quantities that enter the forecast. The main findings and the recommended steps are listed as bullet points in Section~\ref{sec:discussion-conclusion}.

\section{How a selection function error impacts the galaxy density}
\label{sec:sel-func-impact}

In any wide-field cosmological galaxy survey, the definition and characterization of the galaxy sample is of utmost importance to connect the observables back to the cosmological parameters. Fundamentally what an observer does is count the number of galaxies, up to some flux limit, across the sky and tabulate that either in a catalog or a map. To improve efficiencies such as particular redshift ranges, purity, completeness, etc., one may apply additional cuts in the \textit{corrected} color-magnitude space. For a targeted spectroscopic survey like DESI, this can play a critical role of a one-way decision as DESI assigns a non-zero probability of observance only to galaxies that pass the selection cuts. For photometric surveys like LSST, color-flux cuts can enter via complicated non-linear mappings to assign a galaxy to a tomographic bin for analysis.

There are qualitative consequences for errors in these cuts. A targeted approach like DESI may elect to observe a galaxy that should have never been observed. Additionally, a galaxy that should have been observed never gets observed, and such errors could make certain parts of the sky look more overdense or underdense than they actually are. In the case of photometric surveys, while this process can be reversible in theory, in practice the entire analysis machinery that goes from deep-field calibration of the redshift distribution, $n(z)$, all the way to cosmology is expensive to re-run. Therefore, here too, a misspecification could result in a galaxy being assigned the wrong tomographic bin, which also introduces clustering biases. Most importantly, if these selection errors have coherent angular structure, then those structures appear as large-scale structure signal, driving bias in measured clustering.

In the rest of the section, we derive some simple relations to motivate the impact of selection function errors on clustering analysis. Note that while our paper is mostly focused on the issue of Milky Way extinction, in principle any observable that impacts the color-flux selection space can inherit similar issues, e.g., zero-point errors of the filters.

\subsection{First-order model of the imprint}
\label{sec:imprint-model}

We consider the problem at map level, but the argument also applies for any catalog-based clustering analysis. Let us assume that we have observed a set of galaxies with some color-flux limits and are now trying to characterize the observed population in a pixelized space, where each pixel carries an ID number, $x$. An observer first corrects the magnitudes for Galactic extinction with a reddening map, $E_{\rm sel}$, before the cuts are applied.

Let $n_g^{\rm obs}(x)$ be the number of observed galaxies in pixel $x$, $\delta_g(x)$ the true galaxy overdensity, and $f_{\rm pix}(x)$ the fraction of the pixel covered by the survey, estimated from the randoms. The expected counts are
\begin{equation}
    \left< n_g^{\rm obs}(x) \right>
    = \bar{n}\, f_{\rm pix}(x) \left[1 + \delta_g(x)\right] \mathcal{F}(x),
    \label{eq:nobs}
\end{equation}
where the average is over Poisson sampling at fixed $\delta_g$, and $\bar{n}$ is the mean number of galaxies in a fully covered pixel. $\mathcal{F}(x)$ is the \textit{relative selection function}, i.e., the density of selected targets that a uniform universe would produce at $x$, per unit covered area, normalized to unit mean over the footprint. It is distinct from the coverage $f_{\rm pix}$. A survey with no uncorrected systematics has $\mathcal{F} = 1$ everywhere. We assume every known dependence (depth, seeing, stellar density, and the true extinction through photometric noise) has been removed, which leaves the error of the map used at selection. To first order in that error,

\begin{equation}
    \mathcal{F}(x) = 1 + A\, s(x), \qquad
    s(x) = E_{\rm sel}(x) - E_{\rm true}(x), \qquad
    A \equiv \frac{\partial \ln \bar{n}}{\partial s},
    \label{eq:F}
\end{equation}

\noindent where $s(x)$ is the error of the reddening map used at selection and $A$ is the \textit{response} of the sample to it. Equation~\eqref{eq:F} is valid for $|A s| \ll 1$, with the footprint mean of $s$ absorbed into $\bar{n}$. Because the cuts see the map only through the dereddened magnitudes, $A$ equals the fractional change of the selected density when $E_{\rm sel}$ is shifted at fixed photometry, which is how we measure it (Section~\ref{sec:response}). An error $s$ makes the dereddened magnitude in band $b$ brighter by $R_b s$, where $R_b$ is the extinction in band $b$ per unit $E(B-V)$ (Table~6 of \cite{schlafly-finkbeiner-2011}), so by the chain rule

\begin{equation}
    A = \sum_b R_b\, S_b ,
    \label{eq:A-bands}
\end{equation}

\noindent where $S_b$ is the fractional change of the selected density per magnitude of brightening in band $b$ (Appendix~\ref{app:response}). For a sample defined only by a faint limit in band $b$, $S_b = \ln 10 \; d\log_{10} N / dm_b$, with $N$ the cumulative number counts evaluated at the limit (Eqs.~30 and 31 of \cite{huterer-cunha-fang-2013}), and Eq.~\eqref{eq:A-bands} reduces to $A = R_b \ln 10 \; d\log_{10} N / dm_b$ (Eq.~33 of \cite{huterer-cunha-fang-2013}). The response is therefore larger when the limiting band is bluer (where $R_b$ is larger), and when the counts at the limit are steeper. A color cut contributes through the difference of the two coefficients and the density of objects at the cut, so a color-selected sample can respond more strongly than its magnitude limit alone would suggest. This is the same response that sets the traditional magnification bias term \cite{kramsta-magbias-2021, elvin-pool-magbias-2023-des}. In the cosmological setting, lensing brightens galaxies across the cuts, whereas in our case the map error induces the same signal. The differences are that lensing also dilutes the area and moves all bands equally, whereas a map error moves them in the ratio of the $R_b$.

The error $s$ has three origins. First, a map can mis-estimate the Galactic reddening itself, because of how its method converts an observable into reddening or because of its measurement noise (Section~\ref{sec:dust}); for example, \cite{dennis-lee-2025-end-of-fir} find differences of up to $E(B-V) = 0.08$~mag between far-infrared and stellar reddening maps in Ursa Major. Second, a map can contain emission from large-scale structure, such as the cosmic infrared background (CIB) in the map of Schlegel, Finkbeiner and Davis \cite{sfd98} (hereafter SFD98). Third, a map built from stellar colors inherits the zero-point errors of that photometry \cite{rongpu-desi-dustmap-zhou}. We therefore decompose the error term as

\begin{equation}
    s(x) = \epsilon(x) + I(x) + \zeta(x) ,
    \label{eq:s-decomp}
\end{equation}

\noindent where $\epsilon$ is the \textit{Galactic error}, $I$ the extragalactic leakage and $\zeta$ the zero-point error of the map, all in units of $E(B-V)$. Only $I$ correlates with large-scale structure. Which terms are present in the overall error term $s(x)$ depends on how a map is built. $\zeta$ arises only for maps derived from photometric colors, whereas $I$ has been detected in far-infrared maps and in stellar reddening maps alike, and is absent only from the map based on 21\,cm H\,\textsc{i} emission \cite{chiang-menard-2019}. All three enter the selection through the same response $A$.

A zero-point error $\delta m_b(x)$ in the galaxy photometry itself is different. It moves one band, and changes the density by $-S_b\,\delta m_b$ instead of through $A$. When the reddening map is built from the same photometry, $\zeta$ and $\delta m_b$ are correlated, and we return to this in Section~\ref{sec:template-limits}. Finally, Eq.~\eqref{eq:A-bands} takes the coefficients $R_b$ as fixed. If the extinction curve varies over the sky \cite{singh-rv-schlafly-2026}, the error differs from band to band and is no longer described by one field $s$.

\subsection{How to correct for selection function error once galaxies are observed}
\label{sec:correction}

We now consider a selection map that is not built from photometric colors, so that $\zeta = 0$ and $s = \epsilon + I$. From Eq.~\eqref{eq:F}, the observed overdensity, i.e., the overdensity estimated assuming a uniform selection, is to first order
\begin{equation}
    \delta_g^{\rm obs} = \delta_g + A\,(\epsilon + I) ,
    \label{eq:obs}
\end{equation}
where $\epsilon$ is the Galactic part of the error and $I$ the extragalactic part. The two differ in one respect. While $\epsilon$ is independent of the large-scale structure, $I$ is not. Therefore, only $I$ correlates with the galaxies and with the CMB lensing convergence $\kappa$. We also take $\epsilon$ and $I$ to be uncorrelated with each other. Without any correction, the galaxy auto-spectrum and the cross-spectrum with CMB lensing are therefore
\begin{align}
    C_\ell^{gg,{\rm obs}} &= C_\ell^{gg}
        + 2A\,C_\ell^{gI}
        + A^2 \left( C_\ell^{\epsilon\epsilon} + C_\ell^{II} \right) ,
        \label{eq:cgg-obs} \\
    C_\ell^{\kappa g,{\rm obs}} &= C_\ell^{\kappa g} + A\,C_\ell^{\kappa I} .
        \label{eq:ckg-obs}
\end{align}

The observer cannot subtract the imprint $A\,s$ directly because the map error $s$ is not known. What they have instead is a \textit{template}, i.e., a map $T$ of some measurable quantity that is expected to trace $s$, such as the difference between two estimates of the reddening. The correction is then built by analogy with Eq.~\eqref{eq:F}. While the true selection function is $1 + A\,s$, the observer models it as $1 + \beta\,T$, where the known template takes the place of the unknown error and a fitted slope $\beta$ takes the place of the response. Note that $\beta$ is not derived from the cuts; it is found from the data as the least-squares fit of the observed overdensity against the template over the pixels.\footnote{A constant offset in $T$ cannot be distinguished from the mean density, so $T$ here denotes the template with its footprint mean removed. The model can either be divided out of the observed overdensity or kept and carried into the measured spectra, and the two are equivalent to first order \cite{karim2023-galaxy-window-function}.} Removing the model from the data gives, to first order,
\begin{equation}
    \delta_g^{\rm corr} = \delta_g^{\rm obs} - \beta\,T
    = \delta_g + A\,s - \beta\,T .
    \label{eq:corr-linear}
\end{equation}
The correction is therefore exact only if $\beta\,T = A\,s$, which requires the template to trace the map error perfectly and the fitted slope to equal the response.

We emphasize that the two slopes are different quantities. The response $A$ is a property of the sample, measured by shifting the dereddening in the catalog and recounting, and it tells us how large the imprint is per unit of map error but not where on the sky the error lies. The slope $\beta$ depends on both the sample and the template, since it measures how strongly the observed density follows $T$ across the sky, and it equals $A$ only if the template is the map error itself. For an imperfect template the fit returns a different slope because subtracting $A\,T$ would remove the imprint at the cost of adding everything else the template contains. In Section~\ref{sec:leftover} we follow what is left over when either condition fails.

\subsection{Leftover imprint from an imperfect template}
\label{sec:leftover}

The template that traces the error of the selection map most directly is a \textit{reddening-difference template}, i.e., the selection map minus an independent map. In our example, we take this independent map to be a stellar map, so the template is $T = E_{\rm sel} - E_{\rm star}$. Since both maps contain the true reddening, it cancels in the difference, and the template is left with the error of each map, $T = \epsilon + I - n$, where $n$ is the error of the stellar map. For now we consider an idealized stellar map, whose error $n$ is independent of the other fields and which is free of both zero-point errors and extragalactic signal. In a real analysis these issues come back, and we focus on them in Section~\ref{sec:template-limits}. For this template the slope of the fit of $\delta_g^{\rm obs}$ against $T$ is

\begin{equation}
    \beta = A\,\lambda + \gamma ,
    \label{eq:beta-diff}
\end{equation}

\noindent where $\lambda$ is the fraction of the template variance that comes from the error of the selection map and not from the error of the stellar map, so that $0 \le \lambda \le 1$, and $\gamma$ is the slope of the true overdensity on the template, which is nonzero only because the extragalactic part $I$ correlates with the galaxies (Appendix~\ref{app:derivation}).

With this template, the correction of Eq.~\eqref{eq:corr-linear} subtracts $\beta\,(\epsilon + I)$ from an imprint of $A\,(\epsilon + I)$, so that $(A - \beta)(\epsilon + I)$ remains, i.e., a fraction $1 - \beta/A$ of the imprint. At the same time it adds the stellar map error to the density, scaled by $\beta$. The imprint is therefore removed only if $\beta = A$, and Eq.~\eqref{eq:beta-diff} shows the two effects that prevent this. The first is shrinkage. The template contains the error of the
selection map, which the observed density follows with slope $A$, and the error of the stellar map, which the density does not follow at all. Since the fit can only assign one slope to the whole template, it settles on a value between $A$ and zero, weighted by how much of the template variance each part contributes, which gives $A\,\lambda$.

The second is signal removal, where the intrinsic correlation between the galaxies and $I$ raises the fitted slope through $\gamma$, so the correction subtracts real clustering along with the imprint. The fit cannot tell the two apart. The leftover in the lensing cross-spectrum is $(A - \beta)\,C_\ell^{\kappa I}$ (Appendix~\ref{app:derivation}), and since $A - \beta = A(1-\lambda) - \gamma$, its sign depends on which of the two effects is larger.

Eq.~\eqref{eq:beta-diff} assumes that the two maps make unrelated errors. If a part of the stellar map error follows the Galactic error
$\epsilon$ with slope $\rho$, e.g., when both maps misestimate the reddening of the same clouds, the shared part cancels in the template
and the correction removes less of the imprint (Appendix~\ref{app:derivation}).

For the extragalactic term, the forecast considers the largest and the smallest it can be. The reddening-difference template above is built from the selection map and therefore contains $I$, so it removes part of the extragalactic imprint. If instead the selection map contains $I$ but the correction template does not, e.g., when the template is built from a version of the map that has been cleaned of the CIB, the correction cannot touch the terms in $I$, and the full $A\,C_\ell^{\kappa I}$ stays in the lensing cross-spectrum. We call this the \textit{CIB term}. If the selection map itself is free of extragalactic signal, the leftover is additive in the galaxy auto-spectrum only, and we call this the \textit{Galactic term}. Appendix~\ref{app:derivation} gives the corrected spectra for both.

For any correction $X$ we write the result as
\begin{equation}
    \delta_g^{\rm corr} = \delta_g + A\, r_X ,
    \label{eq:residual-field}
\end{equation}
where the \textit{residual field} $r_X$ is what the correction leaves of the map error, in units of reddening. Without a correction $r_X = \epsilon + I$, while for the reddening-difference template $r_X = (1 - \beta/A)(\epsilon + I) + (\beta/A)\, n$, i.e., the part of the map error that survives plus the error of the stellar map that the correction adds. The forecast of Section~\ref{sec:forecast} injects the power spectrum of $r_X$.

One consequence matters for validation. The fit removes everything in the observed density that follows the template, so the corrected
density is uncorrelated with that template by construction, no matter how much of the imprint is still in the map. To measure the leftover
we therefore need a second template whose error is independent of the one used in the fit. We have kept only first-order terms throughout. The higher-order terms, which couple the residual field to the galaxy field, are given in Appendix~\ref{app:second-order}.

\section{Overview of Data}
\label{sec:data}

\subsection{DESI Legacy Imaging Surveys emission-line galaxies}
\label{sec:data-elg}

We use the photometric emission-line galaxy (ELG) sample of \cite{karim2025-elg-cmb-lensing}, selected from the DESI Legacy Imaging Surveys DR9. The average galaxy depths of DR9 in the $g$, $r$ and $z$ bands are $23.60$, $23.07$ and $22.26$ AB magnitudes \cite{karim2025-elg-cmb-lensing}. The sample has about 24 million objects. It is defined by cuts in dereddened magnitudes and colors, followed by a Random Forest classifier trained on DESI spectroscopy to remove low-redshift interlopers and poor targets (Section~\ref{sec:response} lists the cuts). Objects brighter than $g_{\rm fib} = 22.6$ are removed because most of them are quasars. The photometry is dereddened with SFD98. We refer to the footprint of this sample as the DR9 mask, which covers $f_{\rm sky} \approx 0.25$.

Imaging systematics are corrected with the \textsc{sysnet} neural-network regression \cite{mehdi2020-decals7-img-sysnet} on 12 maps, namely the Point Spread Function (PSF) size, the PSF depth and the galaxy depth in $g$, $r$ and $z$, the \textit{Gaia} stellar density, the Sagittarius stream, and one reddening map. Following \cite{karim2025-elg-cmb-lensing} we use five versions of the weights that differ only in the reddening map (Section~\ref{sec:dust}), and the sample without any weights. In all five versions the reddening map is given to the regression as an absolute map and not as a difference of two maps. An absolute map traces the true reddening, so it removes the map error of Section~\ref{sec:imprint-model} only to the extent that the error follows the dust column (Appendix~\ref{app:derivation}). We also note that \textsc{sysnet} is a non-linear regression on many maps, whereas the model of Section~\ref{sec:correction} is linear in a single template; we use the latter to interpret the results and not to describe the pipeline.

\subsection{DESI DR1 emission-line galaxies}
\label{sec:data-dr1}

In Section~\ref{sec:template-limits} we use the public DESI DR1 ELG clustering catalogs and randoms \cite{desi2025-sample-characteristics}. We use the \texttt{ELG\_LOPnotqso} sample, which has about $2.4$ million redshifts at $0.8 < z < 1.6$ over $5924$~deg$^2$ ($f_{\rm sky} \approx 0.14$), and we refer to its footprint as the DR1 footprint. This sample is also selected with SFD98, but with different cuts from the DR9 sample, so its response $A$ need not be the same, and we measure it in Section~\ref{sec:response}.

The imaging weights of this sample come from a \textsc{sysnet} regression on 10 maps \cite{alberto-2024-elg-sysnet}. Unlike our DR9 runs, this regression does not use an absolute reddening map. It uses instead the differences between the two versions of the DESI stellar reddening map and SFD98, $\Delta E(B-V)_{g-r}$ and $\Delta E(B-V)_{r-z}$, together with an H\,\textsc{i} column density map, the stellar density, and the PSF size and galaxy depth in three bands. The regression is trained separately in two redshift bins and in the two imaging regions, and the weights are clipped to $[0.5, 2.0]$. Since the error of SFD98 follows variations of the extinction curve \cite{singh-rv-schlafly-2026}, the two differences are not expected to carry the same information. We form the angular density with the completeness and redshift-failure weights only, which we call the sample without imaging weights, and with the imaging weights in addition. The randoms carry the completeness weights of the catalog, and the two imaging regions are normalized separately.

\subsection{Reddening maps}
\label{sec:dust}

Table~\ref{tab:maps} lists the maps. SFD98 is the map used to select both galaxy samples. The DESI stellar map, which we call the Zhou map, exists in two versions built from $g-r$ and from $r-z$ stellar colors; we use the $g-r$ version throughout, as its authors recommend, and the $r-z$ version for one test in Section~\ref{sec:template-limits}. For the Mudur map, we find that its reddening scale relative to CSFD is not constant across the sky. The slope of the Mudur map against CSFD is about $1.0$ to $1.2$ at declinations between $-20^\circ$ and $+25^\circ$ and drops to $0.40$ to $0.76$ between $32^\circ$ and $65^\circ$. Since a template whose amplitude drifts across the sky could produce spurious structure in a difference map, we also use a version of the Mudur map rescaled by this slope as a function of declination, and we quote results for it unless stated otherwise. The template $E_{\rm SFD98} - E_{\rm Zhou}$ shares more of its power with $E_{\rm SFD98} - E_{\rm Mudur}$ when the Mudur map is scaled by a single factor (Appendix~\ref{app:calibration-tests}). If the drift were an error of SFD98, both differences with the map scaled by a single factor would contain it, and the rescaling would lower the shared power. The rise therefore suggests that the drift belongs to the Mudur map and not to SFD98. The rescaling ties the amplitude of the Mudur map to that of CSFD in each range of declination, so a difference of the two maps cannot show a part of the error of SFD98 that changes its scale with declination. The \textsc{sysnet} run corrected with the Mudur map uses the map as published.

\begin{table}[h]
\centering
\small
\begin{tabular}{l p{4.2cm} l c c}
\hline
Map & Tracer & Resolution & Correction & Measurement \\
\hline
SFD98 \cite{sfd98} & far-infrared emission & $6.1'$ & yes & yes \\
CSFD \cite{csfd} & SFD98, extragalactic signal removed & $6.1'$ & yes & yes \\
\textit{Planck} GNILC \cite{planck-gnilc} & far-infrared emission & variable & yes & yes \\
Zhou \cite{rongpu-desi-dustmap-zhou} & DESI stellar spectra, LS colors & $13.7'$ (pixel) & yes & yes \\
Mudur \cite{nayantara-mudur-dust} & PS1 and 2MASS stellar photometry & $15'$ & yes & yes \\
H\,\textsc{i} \cite{lenz2017-hi-dust} & 21\,cm emission & $16.1'$ & no & yes \\
\hline
\end{tabular}
\caption{Reddening maps used in this paper. ``Correction'' marks the maps given to the \textsc{sysnet} regression as the absolute reddening template of a run on the DR9 sample. ``Measurement'' marks the maps used to build the difference templates of Section~\ref{sec:results-reddening}, which measure what a correction leaves and are never given to the regression. Resolutions are beam widths except for the Zhou map, which is given as the size of its HEALPix pixels.}
\label{tab:maps}
\end{table}

Galactic reddening is mapped with several kinds of tracers, and each converts a different observable into reddening under its own assumptions. Far-infrared (FIR) maps use the thermal emission of dust and convert it to reddening with a temperature correction. SFD98 is built from the IRAS $100\,\mu$m map with a temperature correction from COBE/DIRBE, and is calibrated on the colors of elliptical galaxies \cite{sfd98}. Since the cosmic infrared background (CIB) is also bright at these wavelengths, FIR maps contain emission from large-scale structure. The CIB is emitted over a broad range of redshift \cite{schmidt-cib-redshift-2015}, and the part that enters SFD98 declines with redshift, as expected for a map built at $100\,\mu$m \cite{chiang-menard-2019}. The \textit{Planck} GNILC map \cite{planck-gnilc} separates the CIB from Galactic dust statistically, which reduces this leakage but does not remove it \cite{chiang-menard-2019}, and CSFD \cite{csfd} removes the extragalactic signal from SFD98 using templates built from WISE galaxy density fields.

Stellar maps measure the reddening from the difference between the observed and the intrinsic colors of Milky Way stars. The Zhou map \cite{rongpu-desi-dustmap-zhou} takes the intrinsic colors from DESI spectra and the observed colors from the Legacy Surveys photometry, while the Mudur map \cite{nayantara-mudur-dust} uses Pan-STARRS1 and 2MASS photometry alone. Stellar maps are limited by the number of stars per pixel, which is lowest at the high Galactic latitudes where cosmological surveys observe, and they inherit the zero-point errors of the photometry they use. Note that the Legacy Surveys zero points are themselves tied to Pan-STARRS1 \cite{rongpu-desi-dustmap-zhou}, so the Zhou map, the Mudur map and the galaxy photometry may share part of their calibration errors and are not fully independent.

Finally, the H\,\textsc{i} map \cite{lenz2017-hi-dust} uses the $21$\,cm emission of neutral hydrogen as an indirect tracer, which assumes a constant dust-to-gas ratio and holds only at low column density. Mid-infrared emission from small grains is a fourth tracer \cite{meisner-finkbeiner-mir-pah-dust}, which we do not use here.

These differences are the origin of the error terms of Section~\ref{sec:imprint-model}. For the Galactic error, Lee et al. \cite{dennis-lee-2025-end-of-fir} show that FIR maps under-predict the reddening relative to stellar maps where the abundance of small grains is high, by up to $\Delta E(B-V) = 0.08$~mag in the Ursa Major molecular cloud complex, and Singh et al. \cite{singh-rv-schlafly-2026} find that the error of SFD98 correlates with variations of the extinction curve. For the extragalactic leakage, Chiang \& M\'{e}nard \cite{chiang-menard-2019} cross-correlated dust maps with spectroscopic galaxies and quasars and detected imprints not only in FIR maps but also in the Meisner \& Finkbeiner map \cite{meisner-finkbeiner-mir-pah-dust} and in stellar reddening maps built from Pan-STARRS1 photometry, where they point to imperfect separation of stars from galaxies and quasars. Only the H\,\textsc{i} map showed none.

\begin{figure}[h]
    \centering
    \includegraphics[width=\linewidth]{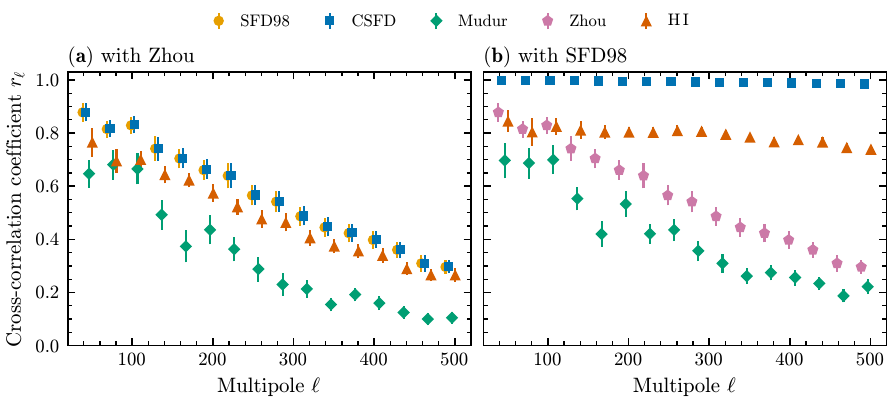}
    \caption{Cross-correlation coefficient of reddening maps with the Zhou map (left) and with SFD98 (right), as a function of $\ell$. All maps are strongly correlated on large scales. The coefficient falls toward small scales, and for the two stellar maps part of this fall is noise. CSFD and SFD98 are almost fully correlated, so the extragalactic signal removed from SFD98 is a small part of its power. The H\,\textsc{i} map and SFD98, both of low noise, correlate at about $0.8$ at all scales. The spectra are computed at \texttt{nside} = 1024 on a mask that is common to all the maps for ease of comparison ($f_{\rm sky} = 0.228$). The mask keeps the pixels where every map has data, with its edges drawn on a grid of \texttt{nside} = 64, and has a C1 apodization of $1^{\circ}$. The error bars come from a delete-one jackknife over 40 equal-area regions (5 Dec $\times$ 8 RA) of the same footprint, re-apodized each time.}
    \label{fig:dust-dust-xcorr}
\end{figure}

Figure~\ref{fig:dust-dust-xcorr} compares the maps on a common mask. The H\,\textsc{i} map and SFD98, which both have low noise, correlate at about $0.8$ at all scales, so two tracers do not carry the same information even where noise is negligible. For the stellar maps the coefficient falls toward small scales. This is what we expect if the error $n$ of the stellar map takes over at small scales, although the figure alone cannot tell how much of the fall comes from noise and how much from a real difference between the maps. The distinction matters at all scales, because the slope of a correction is fitted over pixels (Section~\ref{sec:leftover}), so noise at the pixel scale leaves part of the imprint in on large scales as well. We measure the noise of the stellar map directly in Section~\ref{sec:template-limits}.

Such differences between reddening maps already affect cosmological analyses, as the change in the tension with \textit{Planck} found by Karim et al.~\cite{karim2025-elg-cmb-lensing} shows (Section~\ref{sec:intro}) for the DR9 sample that changing only the absolute reddening map given to the regression changes the tension with \textit{Planck} in $\sigma_8$--$\Omega_m$ from $4.29\sigma$ with SFD98 to $3.07\sigma$ with the Zhou $g-r$ stellar reddening map, with everything else in the analysis kept fixed. The Galactic error also affects the distance moduli of Type Ia supernovae, which are corrected object by object and are therefore sensitive to dust structure that a noise-limited map cannot resolve \cite{elliot2026-snia-dust}. The problem is particularly relevant for southern surveys such as LSST and the Simons Observatory, because no stellar reddening map of comparable precision at high Galactic latitude exists below a declination of $-30^{\circ}$; the all-sky map from \textit{Gaia} \cite{delchambre-gaia-tge-2023} relies on far fewer stars per pixel. Observations of the $3.3\,\mu$m feature of small grains with SPHEREx \cite{crill-spherex-2020} offer a promising way to improve FIR maps \cite{dennis-lee-2025-end-of-fir}.

\subsection{CMB lensing}
\label{sec:data-lensing}

We use the \textit{Planck} PR4 lensing map \cite{planck-pr4} and the ACT DR6 lensing map \cite{mat-madhavacheril-dr6-act-cmblensing, qu-2024-act-cmb-lensing}, the latter in its baseline version and in the version with the CIB deprojected \cite{mccrann-act-cib-deprojected-2024}. A reconstructed lensing map can itself carry residual CIB or Galactic foregrounds, whereas the model of Section~\ref{sec:correction} assumes the true convergence, and we use the CIB-deprojected version to test for this. For the forecast we use the lensing noise expected for the Simons Observatory \cite{so-forecast-2019} (Section~\ref{sec:forecast}).

\section{Results: What have we learned about the DESI selection function and its relationship with reddening?}
\label{sec:results-reddening}

To recap from Eq.~\eqref{eq:F}, the size of the imprint is the product of two things, the response $A$ of the sample and the error $s$ of the reddening map. We take them in turn. Section~\ref{sec:response} measures the response of the DR9 photometric ELG sample, which can be done from the catalog alone. We then turn to the error of the map. Section~\ref{sec:lss-in-maps} measures its extragalactic part, which is the part that enters the cross-spectrum with CMB lensing. Section~\ref{sec:regression-abs-ebv-residual} measures what the standard correction with an absolute reddening map leaves of its Galactic part. Finally, Section~\ref{sec:template-limits} asks how much of the imprint the reddening-difference template of DESI removes for the DR1 spectroscopic sample.

\subsection{The response of the DESI ELG sample is large and can be measured}
\label{sec:response}

\begin{figure}[h]
    \centering
    \includegraphics[width=0.77\linewidth]{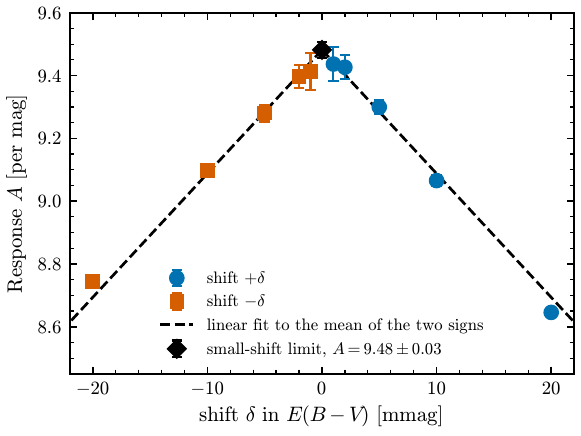}
\caption{Response $A$ of the photometric ELG sample to a shift $\delta$ of the reddening used for dereddening. (Circles and squares) the estimator of Appendix~\ref{app:A-shift-estimator} for positive and negative shifts. The fall with the size of the shift comes from the estimator and not from the sample. (Dashed line) linear fit to the mean of the two signs. (Diamond) the small-shift limit, $A = 9.48 \pm 0.03$ per mag, which we adopt. Error bars are from a delete-one jackknife over 100 regions, and are smaller than the symbols at the larger shifts.}
    \label{fig:response-shift}
\end{figure}

The response $A$ measures how many galaxies enter and leave the sample when their dereddened magnitudes change, so it depends on the selection cuts. Both the photometric and the spectroscopic samples start from the DESI ELG target selection \cite{raichoor-target-selection-elg-2023}, which consists of a faint limit in the $g$-band fiber magnitude, $g_{\rm fib}$, i.e., the magnitude of the flux that falls within a DESI fiber, and a selection box in the $g-r$ versus $r-z$ plane, both applied to dereddened photometry. The box selects galaxies at $0.6 < z < 1.6$ with strong [O\,\textsc{ii}] emission. This selection was developed through dedicated campaigns, initially with MMT Binospec \cite{karim-mmt-binospec-target-selection-elg-2020} and then with the early phase of DESI \cite{raichoor-prelim-elg-target-selection-2020, raichoor-target-selection-elg-2023}. Beyond this common starting point the two samples differ, because they were optimized differently for completeness and purity. The DR1 spectroscopic sample uses the higher-priority part of the box and relies on the spectra to remove quasars and failed redshifts. The DR9 photometric sample has no spectra, so we keep the full box and add cuts that make the sample purer, calibrated on the 20 ELG fields of the DESI Survey Validation 3 phase \cite{karim2025-elg-cmb-lensing}. In particular, we remove objects brighter than $g_{\rm fib} = 22.6$, most of which are quasars, and apply a Random Forest classifier. Table~\ref{tab:color-cuts} in Appendix~\ref{app:color-cuts} summarizes the cuts of both samples. Since the cuts differ, the two samples need not have the same response, and in this subsection we measure it for the photometric sample. 

Since $A$ is the response to an error in the reddening map, we take the photometric catalog and add a constant shift $\delta$ to the reddening used for dereddening, between $\pm 1$ and $\pm 20$ mmag in $E(B-V)$. This makes the dereddened magnitude in band $b$ brighter by $R_b\,\delta$, where $R_b$ is the extinction in band $b$ per unit $E(B-V)$. For the Legacy Surveys photometry, $R_g = 3.21$, $R_r = 2.17$ and $R_z = 1.21$.\footnote{https://www.legacysurvey.org/dr9/catalogs/\#galactic-extinction-coefficients} We then repeat the selection and compare the result with the unshifted catalog. Since $A$ is defined as a derivative, the quantity we want is the limit of these measurements as the size of the shift goes to zero, which we call the small-shift limit. A detailed description of the estimator is given in Appendix~\ref{app:A-shift-estimator}. Figure~\ref{fig:response-shift} shows the result, which is
\begin{equation}
    A = 9.48 \pm 0.03 \ \text{per mag} .
    \label{eq:A-value}
\end{equation}
Here ``per mag'' refers to a magnitude of error in $E(B-V)$, and since $A$ is a logarithmic derivative, it gives the fractional change in the density for a small error, $\Delta n / n \simeq A\,s$, where $s$ is the error of the map. An error of 1 mmag in the reddening map therefore changes the density of this sample by about $0.95\%$, and an error of 10 mmag by about $9.5\%$. For scale, 10 mmag of $E(B-V)$ is only about 32 mmag in the $g$ band. The relation is linear only while $A\,s$ is small, so it should not be extrapolated to errors approaching a magnitude.

\begin{table}[h]
\centering
\small
\begin{tabular}{l c}
\hline
Quantity & Value [per mag] \\
\hline
\multicolumn{2}{l}{\textit{Response to a reddening error}} \\
Small-shift limit, full footprint (adopted) & $9.48 \pm 0.03$ \\
North & $9.50 \pm 0.04$ \\
South & $9.48 \pm 0.03$ \\
SFD98 replaced by CSFD, object by object & $9.47 \pm 0.11$ \\
Contribution of the faint limit $g_{\rm fib} < 24.1$ & $7.45 \pm 0.01$ \\
\hline
\multicolumn{2}{l}{\textit{Response to a shift of one band alone}} \\
$S_g$ & $5.01 \pm 0.01$ \\
$S_r$ & $-3.77 \pm 0.01$ \\
$S_z$ & $1.26 \pm 0.01$ \\
$R_g S_g + R_r S_r + R_z S_z$ & $9.47 \pm 0.03$ \\
\hline
\multicolumn{2}{l}{\textit{For comparison}} \\
DESI DR1 spectroscopic sample, matched objects & $10.80 \pm 0.05$ \\
DESI ELG targets \cite{rongpu-desi-dustmap-zhou} & $\approx 10.5$ \\
\hline
\end{tabular}
\caption{Response of the DR9 photometric ELG sample. All values are small-shift limits, except the contribution of the faint limit, which is for a shift of 10 mmag (Table~\ref{tab:response-cuts}). The single-band responses are for a shift that makes the objects brighter. The extinction coefficients are $R_g = 3.21$, $R_r = 2.17$ and $R_z = 1.21$. Errors are from a delete-one jackknife over 100 regions. The value for the DR1 spectroscopic sample is measured on the 90\% of its objects that match the DR9 photometry (Section~\ref{sec:response}). The value for DESI ELG targets is the mean of the changes reported for shifts of $+10$ and $-10$ mmag, for a selection that differs from ours.}
\label{tab:response-summary}
\end{table}

The response is large mostly because of the faint limit in $g_{\rm fib}$. This limit carries about $80\%$ of the response, while the color cuts carry most of the rest and the classifier reduces it slightly (Table~\ref{tab:response-cuts}). The response to a reddening error is therefore close to that of a magnitude-limited sample in a blue band. For such a sample, Section~\ref{sec:imprint-model} gives $A = R_g \ln 10 \; d\log_{10} N / dm$, where $N$ is the cumulative count of the sample at its limit. Inverting this relation with the contribution of the faint limit (Table~\ref{tab:response-summary}) gives a slope of about $1.0$ per mag. For comparison, a uniform population in Euclidean space has a slope of $0.6$, which bright galaxy counts follow \cite{yasuda2001}, and the counts of faint galaxies are flatter \cite{melcalfe2001-flatter-number-count}. The slope is steeper here because $N$ counts only the selected objects, which lie in a narrow range of magnitude. It is similar to the magnification slope measured for the same sample \cite{karim2025-elg-cmb-lensing}.

Shifting one band at a time gives the responses $S_b$ of Eq.~\eqref{eq:A-bands}, i.e., the fractional change of the selected density per magnitude of brightening in band $b$ alone. Table~\ref{tab:response-summary} lists them. Their sum weighted by the extinction coefficients agrees with Eq.~\eqref{eq:A-value}, so Eq.~\eqref{eq:A-bands} holds for this sample. The single-band responses are larger than one would guess from the reddening response, and of opposite sign in $g$ and $r$, because a shift of one band also moves the colors.

A reddening error moves $g$ and $r$ together, the color-dependent parts largely cancel, and the faint limit dominates. A calibration error behaves differently, because a zero-point error in one band moves that band alone and the colors shift by the full amount. For the same shift in the $g$ band, a zero-point error changes the density by $S_g \approx 5.0$ per mag, whereas a reddening error changes it by $A / R_g \approx 3.0$ per mag.

Two further checks support our findings (Table~\ref{tab:response-summary}). First, we replaced SFD98 by CSFD (large-scale structure-deprojected SFD98) object by object and obtained the same response. Second, the response is the same in the north and in the south and shows no trend with depth or with reddening, which supports the single value of $A$ assumed in Section~\ref{sec:imprint-model}.

For full comparison, we also measure the response of the DR1 spectroscopic sample. Its catalog does not carry the photometry, so we match it to the DR9 photometric catalog by target identifier, and 90\% of the objects match. The others lie either outside the limits of the photometric catalog, e.g., brighter than $g_{\rm fib} = 22.6$, or in parts of the northern imaging where that catalog is incomplete. We hold fiber assignment and spectroscopic success fixed, since a reddening error changes the dereddened magnitudes used for targeting and not the observed flux. On the matched objects we find $A = 10.80 \pm 0.05$ per mag. If the objects outside the limits did not respond at all, the response of the full sample would be lower by at most 0.35. This is close to the response of about 10.5 per mag that Zhou et al.~\cite{rongpu-desi-dustmap-zhou} find for DESI ELG targets, whose selection is close to that of DR1. The response is larger than that of the photometric sample because of one color cut. The line $g-r < -1.2\,(r-z) + 1.3$ of DR1 lies 0.3 mag lower in $g-r$ than its counterpart in the photometric sample, and it passes through a denser part of the color distribution. It carries 29\% of the response, against 11\% for the photometric sample. Two samples drawn from the same target selection therefore differ in response by 10 to 14\%.

\subsection{Detection of a correlation between CMB lensing and the SFD98 and GNILC reddening maps}
\label{sec:lss-in-maps}

\begin{figure}[ht]
    \centering
    \includegraphics[width=\linewidth]{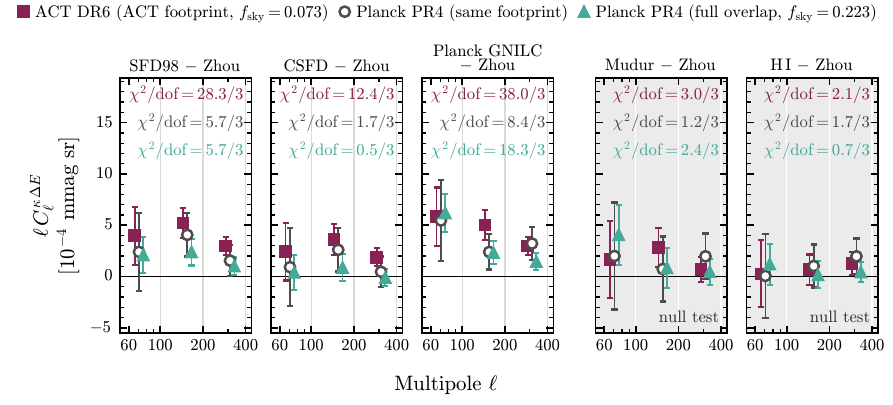}
    \caption{Cross-spectra of CMB lensing with the difference of two reddening maps, $E_X - E_{\rm Zhou}$, for the maps $X$ named in the panels. Three lensing configurations are shown, namely ACT DR6 on its footprint ($f_{\rm sky} = 0.073$), \textit{Planck} PR4 on the same pixels, and \textit{Planck} PR4 on its full overlap with the maps ($f_{\rm sky} = 0.223$). The H\,\textsc{i} map is valid on less of the sky, so its panel uses $f_{\rm sky} = 0.050$ and $0.152$. Errors are from a 100-region jackknife. Each panel gives $\chi^2$ against zero over three bands, written as $\chi^2/{\rm d.o.f.}$ and including the Hartlap factor. The significances quoted in the text and in Table~\ref{tab:kappa-significance} use the Hotelling distribution, so they are lower than a direct conversion of these $\chi^2$ values would give.}
    \label{fig:dust-v-kappa}
\end{figure}

We now turn to the error of the map, starting with its extragalactic part $I$. This is the part that correlates with the large-scale structure, so it is the part that enters the galaxy--lensing cross-spectrum. A key motivation for this paper was to explore how much the CIB leakage in reddening maps could impact galaxy--CMB lensing analyses, as the lensing reconstruction noise keeps getting lower. While the leakage has been shown using SDSS galaxies and quasars \cite{chiang-menard-2019}, we ask the same question with CMB lensing, which does not depend on any galaxy sample and measures directly the term $C_\ell^{\kappa I}$ of Eq.~\eqref{eq:ckg-obs}. Figure~\ref{fig:dust-v-kappa} shows the cross-spectrum of lensing with $E_X - E_{\rm Zhou}$ for several maps $X$. The Galactic content of such a difference does not correlate with the true lensing field, so a signal indicates extragalactic structure in one of the two maps. A reconstructed lensing map can also carry residual foregrounds, and for this reason we include two differences without far-infrared content, H\,\textsc{i} $-$ Zhou and Mudur $-$ Zhou, as null tests.

We test each cross-spectrum against zero with the $\chi^2$ of its three bands in multipole, $[30, 100)$, $[100, 200)$ and $[200, 400)$. The cross-spectra null test consists of fifteen tests for the five differences and the three lensing configurations. The $\chi^2$ distribution assumes that the covariance is known exactly, whereas ours is estimated from a delete-one jackknife over 100 regions and is itself noisy. For a covariance estimated from a finite number of independent samples the statistic follows the Hotelling $T^2$ distribution instead, which is the multivariate analog of using Student's $t$ distribution in place of a Gaussian when a variance is estimated from few samples \cite{sellentin-heavens-2016}. We apply it to the jackknife covariance as an approximation. It has heavier tails than a $\chi^2$ distribution, so the same value of the statistic corresponds to a lower significance, by $0.1$ to $0.5\sigma$ for our largest values. We call the result the local significance.

We also quote a global significance, which accounts for the look-elsewhere effect of having performed fifteen tests and for the dependence between them. For this we simulate the lensing maps and keep the reddening maps as they are. We draw $10^5$ Gaussian realizations of the ACT and \textit{Planck} lensing maps together, with power spectra and a cross-correlation matched to those measured from the real maps. We then cross each realization with the real map differences. These simulations contain no correlation with the reddening maps, so they give the joint distribution of the fifteen statistics when no signal is present. The global significance of a test is obtained from the probability, under this distribution, that the most significant of the fifteen tests reaches the local significance we measure for it.\footnote{We convert each simulated $\chi^2$ to a probability by its rank among the simulations of the same test, so that the simulations provide only the dependence between the tests and each local significance remains the one based on the jackknife.} We verified on the same kind of simulations that the jackknife overestimates the variance of the band powers, by a factor of about $1.6$ in the lowest band and $1.1$ to $1.2$ in the other two, so the significances we quote are conservative.

\begin{table}[ht]
\centering
\small
\begin{tabular}{l l c c c}
\hline
Map difference & Lensing & $\chi^2 (3 ~{\rm d.o.f.})$ & Local & Global \\
\hline
GNILC $-$ Zhou & ACT & $38.0$ & $5.1\sigma$ & $> 4\sigma$ \\
SFD98 $-$ Zhou & ACT & $28.3$ & $4.4\sigma$ & $3.9\sigma$ \\
GNILC $-$ Zhou & \textit{Planck}, full overlap & $18.3$ & $3.4\sigma$ & $2.7\sigma$ \\
GNILC $-$ Zhou & \textit{Planck}, same pixels & $8.4$ & $2.0\sigma$ & $1.0\sigma$ \\
CSFD $-$ Zhou & ACT & $12.4$ & $2.7\sigma$ & $1.8\sigma$ \\
SFD98 $-$ Zhou & \textit{Planck}, both & $5.7$ & $1.5\sigma$ & -- \\
CSFD $-$ Zhou & \textit{Planck}, both & $1.7$, $0.5$ & -- & -- \\
Mudur $-$ Zhou & all three & $3.0$, $1.2$, $2.4$ & -- & -- \\
H\,\textsc{i} $-$ Zhou & all three & $2.1$, $1.7$, $0.7$ & -- & -- \\
\hline
\end{tabular}
\caption{Significance of the cross-spectra of Figure~\ref{fig:dust-v-kappa}. The $\chi^2$ values are those printed in the figure. The local significance uses the Hotelling distribution for a covariance from 100 jackknife regions, and the global significance accounts for the fifteen tests. A dash marks a result that is not significant. Where several configurations are grouped, the $\chi^2$ values are listed in the order ACT, \textit{Planck} on the same pixels, \textit{Planck} on its full overlap.}
\label{tab:kappa-significance}
\end{table}

Two of the differences are detected. Both SFD98 $-$ Zhou and GNILC $-$ Zhou correlate with ACT lensing at about $4\sigma$ after accounting for the fifteen tests. For GNILC the correlation also appears with \textit{Planck} on its full overlap, at lower significance. For SFD98 it does not appear with \textit{Planck} in this test, either on the same pixels or on the full overlap. The GNILC method separates the CIB from Galactic dust using their different angular power spectra, and its authors describe the result as a map with reduced CIB contamination \cite{planck-gnilc}. Our measurement shows that what remains is still detected by CMB lensing, in line with the result of \cite{chiang-menard-2019} from galaxies. Both null tests are consistent with zero in every configuration, so the signals in the other panels do not come from Galactic foregrounds in the lensing maps. The null results also bear on the stellar maps themselves. The H\,\textsc{i} map has been shown to be free of extragalactic signal \cite{chiang-menard-2019}, so the result for H\,\textsc{i} $-$ Zhou means that none is detected in the Zhou map. The result for Mudur $-$ Zhou means that the two stellar maps do not differ in their extragalactic content at this sensitivity.

For CSFD $-$ Zhou the ACT cross-correlation shows an excess at $2.7\sigma$, which falls to $1.8\sigma$ once we account for the look-elsewhere effect. \textit{Planck} shows none, although on its full overlap it would detect a signal of this size at about $3\sigma$. We do not regard this as a detection. It is also not evidence that CSFD is free of extragalactic signal, which the lensing maps of the Simons Observatory will have the sensitivity to test. Our test uses the same ACT and Zhou maps on the same pixels as the test of SFD98 $-$ Zhou, so the two $\chi^2$ values can be compared directly. The $\chi^2$ of CSFD $-$ Zhou is close to half as large, i.e., its amplitude is about two thirds of that of SFD98 $-$ Zhou. If the excess is real, a large part of the extragalactic signal of SFD98 is still present in CSFD, and the term that we measure below from $E_{\rm SFD98} - E_{\rm CSFD}$ is a lower bound.

A more direct statement about SFD98 uses the component that \cite{csfd} removed from it, $I = E_{\rm SFD98} - E_{\rm CSFD}$. Since $I$ was reconstructed from galaxy density fields, its correlation with lensing is expected, and what we add here is the size of the resulting term for a real sample. By Eq.~\eqref{eq:ckg-obs}, a sample selected with SFD98 gains $A\,C_\ell^{\kappa I}$ in its cross-spectrum with lensing, and for the response of the ELG sample this is $5$--$6\%$ of the measured $C_\ell^{\kappa g}$ in each of the three bands, with consistent values from \textit{Planck} PR4, ACT DR6, and the CIB-deprojected ACT DR6 map.\footnote{Measured on the common mask of the three lensing maps ($f_{\rm sky} = 0.052$), with errors from 300 Gaussian simulations of the lensing maps. The errors do not include the $8$--$15\%$ uncertainty of the measured $C_\ell^{\kappa g}$.} This assumes that CSFD removes the whole extragalactic signal. The term is proportional to $A$, and its spectrum is the CIB term that we inject in the forecast (Section~\ref{sec:forecast}).

\subsection{A regression on an absolute reddening map removes part of the Galactic imprint}
\label{sec:regression-abs-ebv-residual}

\begin{figure}[h]
    \centering
    \includegraphics[width=\linewidth]{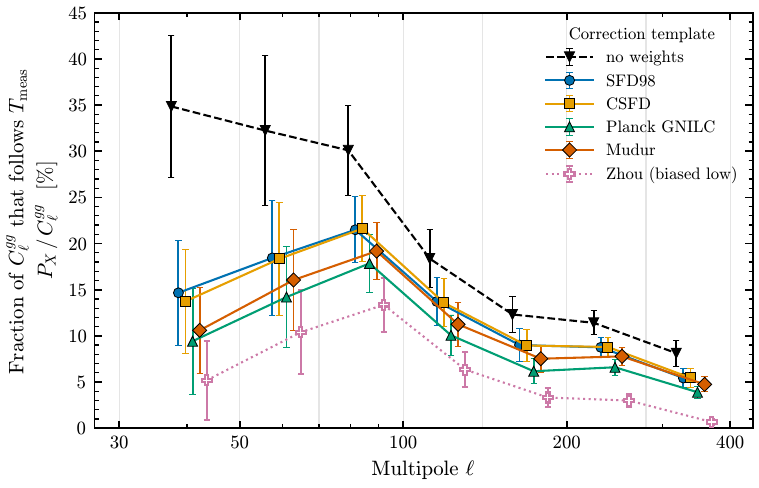}
    \caption{Power in the galaxy density that follows the measurement template $T_{\rm meas} = E_{\rm CSFD} - E_{\rm Zhou}$, as a fraction of the auto-spectrum of each run (Eq.~\ref{eq:PX}). Black shows the sample with no imaging weights. Colors show the \textsc{sysnet} runs, which differ only in the absolute reddening map used as the correction template. A value of $100\%$ would mean that the map consists of imprint only. The run that used the Zhou map as its correction template is biased low, because the same map enters the measurement template. Errors are from a 100-region jackknife.}
    \label{fig:leftover-imprint}
\end{figure}

The rest of the map error is Galactic. It does not correlate with lensing, so we measure it through what it leaves in the galaxy auto-spectrum after the standard correction.

The standard correction uses an \textit{absolute} reddening map, i.e., the map itself is given to the regression as one of its templates. This is designed for a different effect of dust from the one we study. Dust dims the galaxies, which changes the photometric noise and the number of objects detected, so the density depends on the amount of dust on the sky, and a map of that amount is the right template for it. The imprint of Section~\ref{sec:imprint-model} depends instead on the \textit{error} of the map used at selection. The photometric sample is selected once, on photometry dereddened with SFD98, as are the DESI spectroscopic ELGs. The question is not limited to DESI. The object catalog of Rubin Data Preview~2 also lists the SFD98 reddening for each object,\footnote{https://sdm-schemas.lsst.io/dp2.html\#Object.ebv} so a sample selected on photometry dereddened with it will carry the same kind of imprint. We therefore ask how much of that imprint an absolute map removes. As in \cite{karim2025-elg-cmb-lensing}, we run \textsc{sysnet} five times, changing only the reddening map among the templates. The corrected auto-spectrum depends on the reddening map at the lowest multipoles \cite{karim2025-elg-cmb-lensing}, but this spread does not tell us how much of the imprint is left.

To measure what is left we need a template of the map error itself. Two kinds of template are now in play, and we name them to keep them apart. The \textit{correction template} is the map given to the regression. The \textit{measurement template} is used only to measure what a correction leaves, and is never given to the regression. Our measurement template is $T_{\rm meas} = E_{\rm CSFD} - E_{\rm Zhou}$. It contains the Galactic error $\epsilon$ of the selection map, without its extragalactic part, together with the error of the Zhou map. For the density $g_X$ corrected with map $X$, the power that follows $T_{\rm meas}$ is
\begin{equation}
    P_X(\ell) = \frac{\left( C_\ell^{g_X T_{\rm meas}} \right)^2}
                     {C_\ell^{T_{\rm meas} T_{\rm meas}} - N_\ell} ,
    \label{eq:PX}
\end{equation}
where $N_\ell$ is the noise power of the template, taken from the statistical errors of the Zhou map (Appendix~\ref{app:PX-estimator}). $P_X$ is the power spectrum of the part of the density that is proportional to the template. The square of the cross-spectrum carries the units of the template twice and the division by the power of the template removes them, so $P_X$ has the units of the galaxy auto-spectrum and can be compared with it directly. We show it as the fraction $P_X / C_\ell^{g_X g_X}$, which is the squared correlation coefficient between the density and the template, with the template noise removed. If a correction removed the imprint, this fraction would vanish. With no correction, it is the share of the observed auto-spectrum that follows the template.

\begin{table}[ht]
\centering
\small
\begin{tabular}{l c c c}
\hline
 & \multicolumn{2}{c}{Fraction following $T_{\rm meas}$ [\%]} & Second estimate [\%] \\
Correction template & $50 \le \ell < 400$ & $\ell < 100$ & $50 \le \ell < 400$ \\
\hline
No weights & $13.1 \pm 1.4$ & -- & $9.9 \pm 2.4$ \\
SFD98 & $9.4 \pm 1.1$ & $19 \pm 2$ & $6.3 \pm 1.4$ \\
CSFD & $9.4 \pm 1.1$ & $19 \pm 2$ & $6.5 \pm 1.4$ \\
\textit{Planck} GNILC & $6.9 \pm 0.7$ & $15 \pm 2$ & $4.4 \pm 1.1$ \\
Mudur & $8.1 \pm 0.9$ & $17 \pm 2$ & (biased low) \\
Zhou & (biased low) & (biased low) & $4.8 \pm 1.2$ \\
\hline
\end{tabular}
\caption{Fraction of the galaxy auto-spectrum that follows a measurement template, $P_X / C_\ell^{g_X g_X}$, in percent. Each run is divided by its own auto-spectrum. The first two columns use $T_{\rm meas} = E_{\rm CSFD} - E_{\rm Zhou}$ (Eq.~\ref{eq:PX}). The last column uses a second estimate built with $E_{\rm CSFD} - E_{\rm Mudur}$ (see text). An estimate is biased low for the run whose correction template enters its measurement template.}
\label{tab:leftover}
\end{table}

Figure~\ref{fig:leftover-imprint} and Table~\ref{tab:leftover} show the result. Without any weights, about $13\%$ of the auto-spectrum over $50 \le \ell < 400$ follows the template, rising to about $35\%$ in the lowest band. A regression on SFD98 or CSFD lowers this to about $9\%$, and GNILC and the Mudur map do slightly better. The regression therefore removes between a quarter and a half of this power, and leaves the rest.

Two properties of this estimate need care. First, the run that used the Zhou map as its correction template shows the lowest value, but this is expected and says nothing about the correction, since a regression leaves no correlation with its own template (Section~\ref{sec:leftover}). Second, $T_{\rm meas}$ contains more than reddening error. The Zhou map carries the calibration errors of the Legacy Surveys photometry, to which the galaxy density also responds (Section~\ref{sec:response}), so part of $P_X$ comes from calibration and not from the reddening map. As a check, we repeat the measurement with a second stellar map, replacing the cross-spectrum in the numerator of Eq.~\eqref{eq:PX} by the one with $T' = E_{\rm CSFD} - E_{\rm Mudur}$ and the denominator by $C_\ell^{T_{\rm meas} T'}$. This needs no noise model. It is less affected by calibration, although not free of it, since the two stellar maps may share part of their calibration error (Appendix~\ref{app:shared-calibration}). The second estimate gives lower values, but the same conclusion. A regression on SFD98 or CSFD leaves about two thirds of the power it finds without weights.

This result does not mean that the regression fails at its task. It removes the dependence of the density on the amount of dust, and with it the part of the map error that follows the dust column. What it cannot remove is the part of the error that does not follow the column, because an absolute map contains no information about it (Appendix~\ref{app:derivation}). The leftover power measured here is what we inject in the forecast as the Galactic term (Section~\ref{sec:forecast}).

\subsection{The reddening-difference templates of DESI leave part of the Galactic imprint}
\label{sec:template-limits}

The regressions above use absolute reddening maps. The imaging weights of the DESI DR1 ELG sample, which enter its BAO and full-shape analyses, come instead from a regression that includes among its templates the difference between the Zhou stellar map and SFD98, in both its $g-r$ and $r-z$ versions \cite{alberto-2024-elg-sysnet}. We call a template of this kind a reddening-difference template (Section~\ref{sec:leftover}). It traces the error of the selection map directly, and we ask how much of the imprint it removes in practice. We use the public DR1 ELG sample, for which DESI provides the imaging weights that result from this regression (Section~\ref{sec:data-dr1}). From the catalog and its randoms we build two maps of the angular galaxy overdensity, one with the imaging weights applied, which we write $g_w$, and one without them, which we write $g_0$. Throughout this subsection, ``overdensity'' refers to one of these two maps.

To measure what the weights leave, we need a measurement template that was not used to build them (Section~\ref{sec:leftover}). We use $T' = E_{\rm CSFD} - E_{\rm Mudur}$. It shares no stellar map with the DESI templates and, to the extent that CSFD is free of extragalactic signal, it follows the Galactic error only. We cross-correlate $T'$ with $g_w$ and with $g_0$, and take the ratio

\begin{equation}
    R_\ell = \frac{C_\ell^{g_w T'}}{C_\ell^{g_0 T'}} .
    \label{eq:R}
\end{equation}

The ratio is one if the weights leave the imprint along $T'$ untouched and zero if they remove it.

\begin{figure}[ht]
    \centering
    \includegraphics[width=\linewidth]{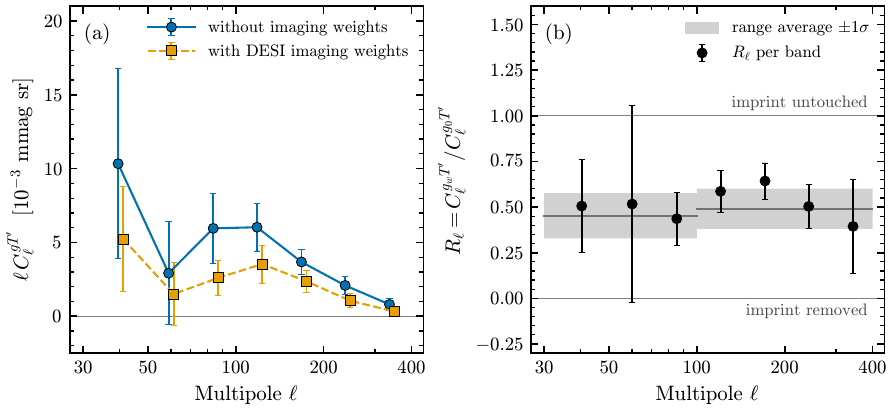}
    \caption{What the DESI DR1 imaging weights leave of the Galactic imprint. (a) Cross-spectrum of the overdensity of the DR1 ELG sample with the measurement template $T' = E_{\rm CSFD} - E_{\rm Mudur}$, without the imaging weights and with them. (b) Their ratio $R_\ell$ (Eq.~\ref{eq:R}). A ratio of one means that the weights leave the imprint along $T'$ untouched, and zero that they remove it. The gray bands show the averages over $30 \le \ell < 100$ and $100 \le \ell < 400$ with their errors. Errors are from a jackknife over the footprint.}
    \label{fig:template-limit}
\end{figure}

\begin{table}[ht]
\centering
\small
\begin{tabular}{l l c}
\hline
Measurement template & Relation to the DESI regression & $R$ \\
\hline
$E_{\rm CSFD} - E_{\rm Mudur}$ & not used & $0.45 \pm 0.12$, $0.49 \pm 0.11$ \\
$E_{\rm CSFD} - E_{\rm HI}$ & H\,\textsc{i} is one of the fitted maps & $0.32 \pm 0.08$ \\
$E_{\rm CSFD} - E_{\rm Zhou}$ & the Zhou map is in the fitted templates & $0.02 \pm 0.04$ \\
\hline
\end{tabular}
\caption{Fraction $R$ of the imprint amplitude that the DESI DR1 ELG weights leave along three measurement templates. For the first row the two values are for $30 \le \ell < 100$ and $100 \le \ell < 400$. The other two rows are from pixel covariances, which include $\ell < 30$, on the part of the footprint where the H\textsc{i} map is valid.}
\label{tab:R-templates}
\end{table}

Figure~\ref{fig:template-limit} and Table~\ref{tab:R-templates} show the result. Along $T'$ the weights leave about half of the imprint amplitude, at all scales we can measure. Along $E_{\rm CSFD} - E_{\rm Zhou}$ they leave nothing. This is the situation described at the end of Section~\ref{sec:leftover}. The Zhou map is in the fitted templates, so $g_w$ cannot correlate with it, and a check against that template alone would have suggested a complete correction.

We cannot establish that $T'$ is a perfect template of the Galactic error. The Mudur and Zhou maps are built from different stars with different methods. However, the Legacy Surveys photometry is calibrated on Pan-STARRS1 \cite{rongpu-desi-dustmap-zhou}, on which the Mudur map is also based. We also find that the two stellar maps agree with each other more closely than either agrees with the H\,\textsc{i} map (Appendix~\ref{app:calibration-tests}). If part of what they share is not the error of SFD98, the ratio along $T'$ is not exactly the fraction of the reddening imprint that is left in $g_w$. The ratio depends little on how the Mudur map is put on the scale of CSFD. With a single scale factor in place of the rescaling in declination (Section~\ref{sec:dust}), it is $0.56 \pm 0.19$ at $30 \le \ell < 100$ and $0.50 \pm 0.11$ at $100 \le \ell < 400$. Both values agree with those of the rescaled map within $1\sigma$, once the correlation between the two estimates is taken into account. Repeating the test with the H\,\textsc{i} map gives about a third, although the H\,\textsc{i} column density is itself one of the maps in the DESI regression. What we can state is that $g_w$ still correlates with templates of the reddening error other than the fitted one, with a third to a half of the amplitude it had before. Therefore, finding no correlation with a single template is not sufficient evidence that a correction has worked.
What limits the correction procedure is the content of the template. A reddening-difference template contains two things, the error of SFD98, which we want, and the error of the Zhou map, which we do not. To see how much of each it holds, we build the same difference with the Mudur map and ask how much the two differences have in common. Over all scales they share about a quarter of their variance. The shared fraction is about a half at the largest scales and falls below $0.2$ at the smallest (Appendix~\ref{app:calibration-tests}). So at least half of the template is not the error of SFD98; this is especially true at small scales, where most of the template is something else.

The rest belongs to the Zhou map, and it has two parts. The first is noise, because each pixel holds a limited number of stars. Noise makes the fitted slope too small, so part of the imprint stays in. This is the shrinkage of Section~\ref{sec:leftover}, and it explains why the shared fraction falls toward small scales. The second is photometric calibration. The Zhou map is built from Legacy Surveys colors, so it carries the calibration errors of that photometry, and the number of selected galaxies responds to the same errors (Section~\ref{sec:response}).

The evidence for the second part comes from a simple experiment. We take $g_0$ and fit it against the single template $E_{\rm SFD98} - E_{\rm Zhou}$ over the pixels of the footprint, with one slope, as in Eq.~\eqref{eq:corr-linear}. The slope tells us by how much the overdensity changes per magnitude.

We can predict this slope if the part of the template that belongs to the Zhou map were noise alone. In that case only one part of the template would matter to the overdensity, the part that is error of SFD98. The overdensity would follow that part with the full response of the sample, and it would ignore the rest. The fitted slope would then be the response multiplied by the fraction of the template that is error of SFD98, which we found above to be about a quarter. The DR1 sample has a response of $10.8$ per mag (Section~\ref{sec:response}), so we would expect a slope of about $10.8 \times 0.25 \approx 2.7$ per mag.

The fit gives $6.3 \pm 0.3$ per mag, more than twice this value. Noise in the Zhou map cannot produce this excess, but three other effects can. The first is the intrinsic correlation of the galaxies with the extragalactic signal in SFD98, i.e., the term $\gamma$ of Eq.~\eqref{eq:beta-diff}. The second is a dependence of $g_0$ on depth, seeing or stellar density, since $g_0$ carries no imaging weights. The third is calibration, since the galaxies and the Zhou map are measured with the same photometry. This fit alone cannot separate them.

A second test we ran points to calibration (Appendix~\ref{app:calibration-tests}). The two versions of the Zhou map measure the same dust through different colors. If the extinction curve is the same everywhere, the dust cancels in their difference, and what remains comes from the photometry. This difference contains no SFD98, so the first effect is absent. \cite{rongpu-desi-dustmap-zhou} show that most of this difference is removed by a correction of the photometric zero points. The overdensity of our DR9 photometric sample, without imaging weights, follows this difference, with the slope expected from the single-band responses of the sample and the published zero-point errors. The signal is also present on the cleanest sky, where a variation of the extinction curve would have little effect, so it is a property of the imaging and not of the dust, most likely calibration.

The single-slope fit also tells us where the limit of the correction lies. With a slope of $6.3$ per mag against a response of $10.8$, a single-slope correction would remove about $60\%$ of the imprint and leave about $0.4$ of its amplitude. This is close to what the full DESI weights leave along $T'$. The two numbers are not exactly the same quantity, since the fitted template contains the extragalactic error of SFD98 and $T'$ does not. With that caveat, a simple linear fit on the template already does most of what the neural network does for the reddening error, which points to the template and not to the regression method as the limit.

A reddening-difference template is therefore only as good as the stellar map in it. The noise of that map weakens the correction. Its calibration adds a second field to which the sample responds differently, and one template cannot serve both. Lower noise in the stellar map would help with the first. The second needs either a calibration template next to the reddening template, e.g., zero points from \textit{Gaia} spectrophotometry \cite{rongpu-desi-dustmap-zhou}, or a stellar map built on independent photometry. Two limits of our measurements should be kept in mind. Every estimate of the leftover in this paper uses a template without extragalactic signal, so the terms that involve $I$ are not measured after correction, and in the forecast we use their uncorrected size. Our statements in this subsection concern the angular overdensity of the DR1 ELG sample. We do not assess the effect on DESI's clustering measurements, and leave the impact of these imprints on the BAO scale and on three-dimensional clustering to future work. 

\section{Forecast setup: a Rubin $\times$ Simons Observatory analysis}
\label{sec:forecast}

Section~\ref{sec:results-reddening} measured what a reddening error does to a real sample. These measurements are its response $A$, i.e., the fractional change in density per magnitude of reddening error, the extragalactic term $A\,C_\ell^{\kappa I}$ that SFD98 adds to the cross-spectrum with CMB lensing, the power $P_X$ that is left in the auto-spectrum after a regression on a reddening map, and the fraction $R$ of the imprint amplitude that the DESI weights leave.

We now ask what these would do to a Stage-IV photometric analysis if they were present in its galaxy sample.
We consider the combination of galaxy clustering from the Rubin Observatory Legacy Survey of Space and Time (LSST) with CMB lensing from the Simons Observatory (SO), which is expected to give some of the tightest constraints on primordial non-Gaussianity and on the growth of structure \cite{so-forecast-2019, schmittfull-seljak-variance-cancel-2018, fang2021-lsst-so-forecast-cmb}. For the galaxy samples we follow the LSST DESC Science Requirements Document (SRD) \cite{lsst-desc-srd}, whose own forecasts do not model Galactic extinction. Table~\ref{tab:forecast-setup} summarizes the setup, and Section~\ref{sec:imprint-injection} describes how the measured terms are injected.

\subsection{Samples and surveys}
\label{sec:forecast-setup}

\begin{figure}
    \centering
    \includegraphics[width=\linewidth]{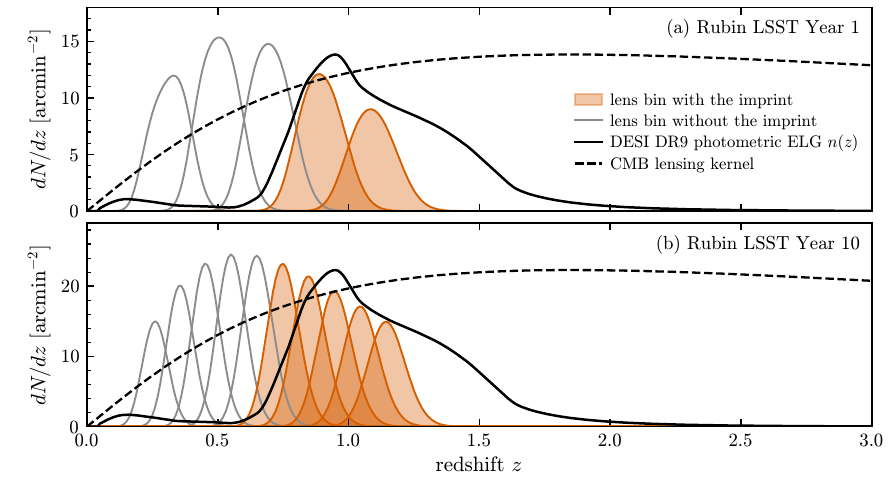}
    \caption{Redshift distributions entering the forecast. Gray and orange curves show the Rubin LSST lens bins for Year~1 (a, five bins) and Year~10 (b, ten bins). The orange, filled bins carry the imprint in the
    forecast. The black solid curve is the redshift distribution of the DESI
    DR9 photometric ELG sample in which the imprint was measured \cite{karim2025-elg-cmb-lensing}, and the black dashed curve is the CMB lensing kernel. The two black curves are scaled to the same peak height for display and
    are not in the units of the vertical axis. The ELG sample has about
    $0.5$~arcmin$^{-2}$, compared to $18$ and $48$~arcmin$^{-2}$ for the lens
    samples.}
    \label{fig:lsst-so-nz}
\end{figure}

\begin{table}[ht]
\centering
\small
\begin{tabular}{p{5.6cm} p{4.2cm} p{4.2cm}}
\hline
 & Year 1 & Year 10 \\
\hline
Lens bins in photometric redshift, $0.2 < z < 1.2$ & 5, width $0.2$ & 10, width $0.1$ \\
Number density [arcmin$^{-2}$] & 18 & 48 \\
Photometric-redshift scatter & $0.03\,(1+z)$ & $0.03\,(1+z)$ \\
Bins with the imprint & 4--5 ($0.8 < z < 1.2$) & 6--10 ($0.7 < z < 1.2$) \\
Sky fraction & $0.30$ & $0.40$ \\
\hline
CMB lensing noise & \multicolumn{2}{p{8.8cm}}{SO forecast at $\ell \ge 30$; \textit{Planck} PR4 level at $8 \le \ell < 30$} \\
Cosmology and spectra & \multicolumn{2}{p{8.8cm}}{\textit{Planck} 2018 \cite{planck2018-cosmo-pararms}; \textsc{camb} \cite{Lewis:1999bs,Lewis:2026mif}; Gaussian covariance} \\
\hline
\end{tabular}
\caption{Setup of the forecast. The lens samples follow the SRD \cite{lsst-desc-srd}.}
\label{tab:forecast-setup}
\end{table}

For our forecast, we only consider the lens samples. They are expected to have
better-known redshift distributions than the sources \cite{fang2021-lsst-so-forecast-cmb}, and their tomographic bins are narrower, allowing us to localize cross-correlation signals better.

The lens samples are those of the SRD for Year~1 and Year~10 (Table~\ref{tab:forecast-setup}, Figure~\ref{fig:lsst-so-nz}). For CMB lensing
we use the reconstruction noise forecast for SO with the baseline sensitivity and the iterative estimator\footnote{\url{https://github.com/simonsobs/so_noise_models}, version 3.1.1,
file \path{nlkk_v3_1_0deproj0_SENS1_fsky0p4_it_lT30-3000_lP30-5000.dat}} at
$\ell \ge 30$. Below that we use a noise level representative of \textit{Planck}
PR4, which enters only the forecast for $f_{\rm NL}$ (Table~\ref{tab:forecast-vectors}). We use the same noise for both years. SO
surveys $40\%$ of the sky, so in Year~1 the overlap is set by the area of Rubin.

The SRD forecast for this sample uses the galaxy auto-spectra only, and we add
the cross-spectrum of each bin with CMB lensing. We depart from the SRD in two
ways. We take a sky fraction of $0.40$ in Year~10, the area of the SO forecast,
where the SRD uses $0.35$; this follows \cite{fang2021-lsst-so-forecast-cmb}.
We also use a more conservative scale cut, with $k_{\rm max}$ lower than its
$0.3\ h\,{\rm Mpc}^{-1}$ (Table~\ref{tab:lmax}).

\paragraph{The bins with the imprint.}
We inject the imprint in the high-redshift lens bins, bins 4--5 in Year~1 and
bins 6--10 in Year~10. As Figure~\ref{fig:lsst-so-nz} shows, these overlap the
peak of the redshift distribution of the ELG sample in which the imprint was
measured. Similar to the DESI galaxies, the LSST lens sample and its
tomographic assignments will use magnitude and color information corrected for
Galactic reddening. As we pointed out in
Section~\ref{sec:regression-abs-ebv-residual}, the Rubin object catalog
currently provides the SFD98 reddening for each object.

Our main assumption is that since LSST will go fainter and observe the same
blue galaxies as the DESI ELGs, the response of these bins could be similar to
that of the DESI photometric ELGs, once tomographic assignments are taken into
account. We therefore give these bins the response of the ELG sample,
Eq.~\eqref{eq:A-value}. Depending on whether the actual response $A$ is
stronger or weaker in the LSST lens sample, the results of our paper will
correspondingly be more severe or less severe, and
Section~\ref{sec:requirement} gives them as a function of the response.

Note that the lens sample as a whole is limited in the $i$ band, where the
extinction coefficient is $R_i = 1.68$ \cite{schlafly-finkbeiner-2011} and the
SRD counts have a slope of $0.359$ per mag (Equation 9 of
\cite{lsst-desc-srd}), so the magnitude-limit result of
Section~\ref{sec:imprint-model} gives a response of only about $1.4$ per mag.
However, we do not know the response of an individual tomographic bin.
Galaxies are assigned to bins by their dereddened colors, so a map error also
moves them across bin edges. This needs to be carefully studied and
characterized.

The forecast for $f_{\rm NL}$ uses only the bins with the imprint, since the low-redshift bins have a bias close to one and a small volume and add little to this measurement. The forecasts for $\sigma_8$, $\Omega_m$ and $\Sigma m_\nu$ use all bins, and we leave the low-redshift bins without an imprint. With a response of $1.4$ these bins would carry about $2\%$ of the power that the ELG response gives in the auto-spectrum and $15\%$ of the term in $C_\ell^{\kappa g}$. This is an optimistic choice, since the response of the low-redshift bins is not known either. Our scenario therefore describes a case where the response is much smaller at low redshift.

\subsection{Data vectors and scale cuts}
\label{sec:forecast-vectors}

\begin{table}[ht]
\centering
\small
\begin{tabular}{l p{4.8cm} p{4.8cm}}
\hline
 & $f_{\rm NL}$ & $\sigma_8$, $\Omega_m$, $\Sigma m_\nu$ \\
\hline
Lens bins & bins with the imprint only & all bins, imprint in the high-redshift bins \\
Galaxy auto-spectra & $30 \le \ell \le \ell_{\rm max}$ & $30 \le \ell \le \ell_{\rm max}$ \\
Cross-spectra with lensing & $8 \le \ell \le \ell_{\rm max}$; also used alone & $30 \le \ell \le \ell_{\rm max}$ \\
Projection & no Limber approximation, with redshift-space distortions & Limber approximation \\
Free cosmological parameters & $f_{\rm NL}$ & $\sigma_8$, $\Omega_m$, $h$; plus $\Sigma m_\nu$ for the neutrino mass \\
Fixed & all others & $n_s$, $\omega_b$; $\Sigma m_\nu = 0.06$~eV when not free \\
Priors on cosmology & none & $\sigma(h) = 0.0054$; for $\Sigma m_\nu$ also $\sigma(\ln A_s) = 0.014$ and $\sigma(\omega_c) = 0.0012$ \\
\hline
Linear bias & \multicolumn{2}{l}{one per bin, no prior} \\
Shift of $n(z)$ & one per bin; $\sigma = 0.0105$, $0.00645$ & one per bin; $\sigma = 0.002$, $0.001$ \\
Width of $n(z)$ & one per bin; $\sigma = 0.21$, $0.0645$ & one for all bins; prior varied \\
\hline
\end{tabular}
\caption{Data vectors and parameters of the two forecasts. All priors are
Gaussian. Where two values are given they are for Year~1 and Year~10. The
priors on cosmology have the widths of the \textit{Planck} 2018 constraints
\cite{planck2018-cosmo-pararms}.}
\label{tab:forecast-vectors}
\end{table}

We run two forecasts, which differ in the data vector and in the parameters that
are free (Table~\ref{tab:forecast-vectors}). In both, the data vector consists
of the auto-spectrum of each lens bin and its cross-spectrum with CMB lensing.

For $f_{\rm NL}$ the signal is on the largest scales, so we compute the spectra
without the Limber approximation and include redshift-space distortions
(Appendix~\ref{sec:theory-fnl}). We use the cross-spectra with lensing down to
$\ell = 8$ and the galaxy auto-spectra from $\ell = 30$. A higher minimum
multipole for the auto-spectrum follows published cross-correlation analyses of
$f_{\rm NL}$, which limit or drop the auto-spectrum on the largest scales
because of imaging systematics
\cite{bermejo2025-dr9-lrg-planck-cmb-lensing, krolewski2024-qso-dr9-planck}.
It is also the lowest multipole at which the leftover power was measured. We
call this the full data vector.

We also consider the more conservative case where only the cross-spectra with
lensing are used. The same analyses adopt it since the cross-spectrum is
expected to be less sensitive to imaging systematics than the auto-spectrum.
In this case the Galactic term does not enter, and the imprint acts only
through the CIB term in $C_\ell^{\kappa g}$.

For $\sigma_8$, $\Omega_m$ and $\Sigma m_\nu$ we use all lens bins in the Limber approximation (Appendix~\ref{sec:theory-amplitude}). The neutrino mass is carried by one massive eigenstate, and the galaxies trace the field of cold dark matter and baryons.

Each bin has a linear bias and a shift and a width of its redshift
distribution as nuisance parameters. For $f_{\rm NL}$ we give the shift and the
width loose Gaussian priors based on the SRD requirements for the clustering
analysis. For $\sigma_8$, $\Omega_m$ and $\Sigma m_\nu$ the result depends on
how well the width is known. We therefore use one width parameter for all bins,
as in \cite{fang2021-lsst-so-forecast-cmb}. We keep their priors on the shifts
and vary the prior on the width (Section~\ref{sec:requirement}).

\begin{table}[ht]
\centering
\small
\begin{tabular}{c c c c}
\hline
$k_{\rm max}$ [$h\,{\rm Mpc}^{-1}$] & Year 1, bin 4 & Year 1, bin 5 & Year 10, bins 6 to 10 \\
\hline
$0.10$ & 211 & 244 & 184 to 253 \\
$0.15$ & 316 & 366 & 276 to 379 \\
$0.20$ & 399 & 399 & 369 to 399 \\
\hline
\end{tabular}
\caption{Maximum multipole $\ell_{\rm max}$ of the bins with the imprint, for the three values of $k_{\rm max}$. For Year~10 the range is over the five bins.}
\label{tab:lmax}
\end{table}

The scale cut of each bin $i$ is $\ell_{\rm max} = \min\left[ k_{\rm max}\, \chi(\bar{z}_i),\, 399 \right]$, where $\chi(\bar{z}_i)$ is the comoving distance to the mean redshift of the bin, and we use three values of $k_{\rm max}$ (Table~\ref{tab:lmax}). The cap at $\ell = 399$ is the highest multipole at which the leftover power was measured.

\subsection{Injecting the imprint and computing the bias}
\label{sec:imprint-injection}

We now describe how the leftover signals that we model and measure in Section~\ref{sec:results-reddening}
enter the forecast. We consider both the case of a Galactic error only and the case of a Galactic error with leakage of the large-scale structure. In the first case, only the galaxy auto-spectrum is affected. As Galactic dust does not know about the large-scale structure, the Galactic error term does not enter the cross-spectrum with CMB lensing. The second is present only if the sample is selected with a map that contains extragalactic signal,
such as SFD98. As this signal traces the large-scale structure, it adds to both the auto-spectrum and the cross-spectrum with lensing.

\paragraph{From the ELG sample to a Rubin bin.}
Consider a lens bin $i$ with a response $A_i$. The galaxies of this tomographic bin are selected using their dereddened color and magnitude information. As an observer does not know the true reddening field, they pick a reddening map that has some error. They now want to correct for this error, e.g., with a regression on the absolute reddening map of SFD98; we denote this correction as $X$ (Table~\ref{tab:leftover}). We also assume that the properties of the error
are the same at any part of the sky as those we measure on the DESI footprint. Therefore, the only quantity that changes from the ELG sample to this bin is the response $A_i$.

Now if the observer applies the correction $X$ to remove the error from the dereddening step, they expect a corrected galaxy overdensity field that matches the true overdensity field. However, the template used in the correction may have errors of its own, or it may not trace the error of the selection map. As a consequence, some imprint is still left in the \textit{corrected} galaxy
overdensity field. By Eq.~\eqref{eq:residual-field} the corrected overdensity
of the bin is
\begin{equation}
    \delta_{g,X}^{{\rm corr},\,i} = \delta_g^i + A_i\, r_X ,
    \label{eq:residual-bin}
\end{equation}
where $\delta_g^i$ is the true overdensity and $r_X$ is the residual field, i.e., the part of the map error in magnitudes that the correction $X$ does not remove. We can now use Eq.~\eqref{eq:residual-bin} to model the leftover signal expected in the presence of this kind of imperfect correction step. The quantity of interest is the departure of each angular power spectrum from its true value at every multipole, which we denote as $\Delta C_\ell$.

\paragraph{Galactic term.}
We first take a selection map whose error is purely Galactic, i.e., it has no extragalactic signal. The residual field $r_X$ is then a pattern on the sky set
by the Milky Way, and it has nothing to do with the large-scale structure. This has two consequences.

The first is that the cross-spectrum of the bin with CMB lensing does not
change. CMB lensing traces the large-scale structure and does not correlate
with a Galactic pattern.

The second is that the auto-spectrum of the bin gains power. In
Eq.~\eqref{eq:residual-bin} the term $A_i\, r_X$ is not correlated with the
true overdensity, so its power simply adds to the true spectrum. This extra
power is $A_i^2$ times the power spectrum of the residual field,
$C_\ell^{r_X r_X}$. We do not measure the residual field itself. What we
measure is the power it leaves in the auto-spectrum of the ELG sample, $P_X$
(Section~\ref{sec:regression-abs-ebv-residual}), which by the same argument is
$A_{\rm ELG}^2\, C_\ell^{r_X r_X}$ with $A_{\rm ELG}$ the response of that
sample, Eq.~\eqref{eq:A-value}. Dividing $P_X$ by $A_{\rm ELG}^2$ therefore
gives the power of the residual field, and multiplying by $A_i^2$ gives the
leftover power in the bin,
\begin{equation}
    \Delta C_\ell^{g_i g_i} = A_i^2\, C_\ell^{r_X r_X}
    = \frac{A_i^2}{A_{\rm ELG}^2}\, P_X(\ell) ,
    \qquad
    \Delta C_\ell^{\kappa g_i} = 0 .
    \label{eq:dCgg}
\end{equation}

Therefore, a sample that is half as sensitive to reddening as the ELGs carries a quarter of their leftover power. Note that this power is additive.
It does not depend on how strongly the galaxies cluster, which is why we carry over the power $P_X$ itself and not a ratio of corrected to true spectra. In our forecast the bins with the imprint have $A_i = A_{\rm ELG}$ (Section~\ref{sec:forecast-setup}), so the ratio in Eq.~\eqref{eq:dCgg} is one. We keep it general because Section~\ref{sec:requirement} varies the response.

Two properties of the measured $P_X$ carry over to the forecast. We measure it with the measurement template $E_{\rm CSFD} - E_{\rm Zhou}$, which we take to contain no extragalactic signal. It also includes a part from the calibration of the Legacy Surveys photometry (Section~\ref{sec:template-limits}), which belongs to that imaging and not to the reddening map. We cannot separate this part, so it is included in the Galactic term that we inject.

Each correction $X$ in Table~\ref{tab:leftover} has its own $P_X$ and gives one
entry of the forecast. We add one more entry for the kind of correction that
DESI applies to its DR1 clustering catalogs. There, each galaxy receives a
weight from a regression of the observed density on a set of imaging templates
\cite{alberto-2024-elg-sysnet}. Two of these are the reddening-difference
templates of \cite{rongpu-desi-dustmap-zhou}. Such a template is meant to trace the error of
the selection map directly, which an absolute reddening map cannot do
(Section~\ref{sec:correction}). This is the most complete correction for the
imprint in use today, so we include it as its own entry.

For this correction we did not measure the leftover power directly. We measured
the fraction $R$ of the imprint that remains in the density field after the
weights are applied. To do so we compared how strongly the weighted and the
unweighted DR1 ELG densities correlate with a template that was not used in the
regression. A perfect correction would give $R = 0$, and a correction that
removes nothing would give $R = 1$. We found $R$ to lie between a third and a
half (Section~\ref{sec:template-limits}). If a fraction $R$ of the imprint
remains, the leftover field is $R$ times the imprint of the uncorrected sample.
Since power scales as the square of the amplitude, the leftover power is
\begin{equation}
    \Delta C_\ell^{g_i g_i} = \frac{A_i^2}{A_{\rm ELG}^2}\, R^2\, P_{\rm none}(\ell) ,
    \label{eq:dCgg-R}
\end{equation}
where $P_{\rm none}$ is the power that the imprint leaves in the ELG sample when no correction is applied (Table~\ref{tab:leftover}). We use the two ends of the measured range, $R = 0.33$ and $R = 0.50$. The lower end is optimistic, since the H\,\textsc{i} map is one of the maps fitted in the DESI regression (Section~\ref{sec:template-limits}). Eq.~\eqref{eq:dCgg-R} also leaves out one term. A correction with a reddening-difference template subtracts the noise of the stellar map along with the imprint, which adds $\beta^2 N_\ell$ to the auto-spectrum (Eq.~\eqref{eq:app-cgg-diff}). We measure $R$ along a template that is independent of the Zhou map, so $R$ does not include this term, and for a given $R$ the Galactic term of this entry is a lower bound. The noise of the Zhou map is white, so it is a growing share of the power of the template toward high multipoles (Appendix~\ref{app:PX-estimator}). It therefore matters least for $f_{\rm NL}$, whose shift follows the leftover power at low multipoles (Section~\ref{sec:results-fnl}). It can matter more for analyses that draw information from small scales, e.g., on the neutrino mass or on modified gravity through a scale-dependent growth of structure, or on $\sigma_8$ from nonlinear scales. Since this term is white, it acts as an extra shot noise of the sample that comes from the stellar map and not from the galaxies. An analysis that fixes the galaxy shot noise to $1/\bar n$, as ours does, reads it as clustering.

\paragraph{Galactic and CIB terms.}
We now take a sample that is selected with SFD98. As we showed in
Section~\ref{sec:lss-in-maps}, this map contains an extragalactic signal $I$ that follows the large-scale structure. The density of the sample then carries the imprint $A_i\, I$ in addition to the Galactic one. We take this part of the imprint to be untouched by the correction, so the corrected overdensity of

Eq.~\eqref{eq:residual-bin} becomes
\begin{equation}
    \delta_{g,X}^{{\rm corr},\,i} = \delta_g^i  + A_i\, r_X + A_i\, I .
    \label{eq:residual-bin-cib}
\end{equation}

Unlike the Galactic pattern, $I$ is correlated with everything that traces the large-scale structure. This gives three new terms.

The first is in the cross-spectrum with CMB lensing. Lensing and $I$ trace the
same structure, so the imprint correlates with the lensing map on its own,
whether or not the galaxies of the bin do. This adds $A_i\, C_\ell^{\kappa I}$,
where $C_\ell^{\kappa I}$ is the cross-spectrum of the lensing map with the
extragalactic signal. An analysis that does not model this term reads it as
extra clustering of the galaxies.

The second and third are in the auto-spectrum. The galaxies of the bin also
trace the large-scale structure, so the imprint correlates with the true
overdensity. This adds $2 A_i\, C_\ell^{\delta_g I}$, where the factor of two
appears because the product of the two terms enters the auto-spectrum twice.
The imprint also has power of its own, $A_i^2\, C_\ell^{II}$. Together,
\begin{align}
    \Delta C_\ell^{\kappa g_i} &= A_i\, C_\ell^{\kappa I} ,
        \label{eq:dCkg-cib} \\
    \Delta C_\ell^{g_i g_i} &= 2 A_i\, C_\ell^{\delta_g I} + A_i^2\, C_\ell^{II} ,
        \label{eq:dCgg-cib}
\end{align}
and these add to the Galactic term of Eq.~\eqref{eq:dCgg}. Note that two of
these terms are linear in the response, while the Galactic term is quadratic.
For a sample with a small response the extragalactic signal therefore matters
more than the Galactic error.

We estimate the extragalactic signal as $\hat I = E_{\rm SFD98} - E_{\rm CSFD}$, the correction that CSFD applies to SFD98. From this map we measure $C_\ell^{II}$ and, with the \textit{Planck} PR4 lensing map,
$C_\ell^{\kappa I}$. We use \textit{Planck} PR4 because its lensing map covers the full footprint of the ELG sample, on which the other spectra are also measured. With ACT DR6, which covers about a third of that footprint, the significance is similar and the amplitude agrees within $1\sigma$. We use the auto-spectrum of $\hat I$ for $C_\ell^{II}$ without correcting for the error of the CSFD estimate, and all the spectra that involve $\hat I$ contain only the part of the extragalactic signal that CSFD removes. If CSFD retains part of this signal (Section~\ref{sec:lss-in-maps}), the three CIB terms are larger than the ones we inject.

We estimate the cross-spectrum $C_\ell^{\delta_g I}$ of Eq.~\eqref{eq:dCgg-cib} from the raw, uncorrected galaxy overdensity of the ELG
sample. The cross-spectrum of this field with the extragalactic signal, $C_\ell^{g_{\rm raw} I}$, has two parts. The first is the one we are after. The galaxies and the extragalactic signal trace the same large-scale structure, so
the true overdensity correlates with $I$. The second comes from the imprint.
The ELG sample was selected with SFD98, so its raw overdensity contains the
term $A_{\rm ELG}\, I$, the same term as the third one in Eq.~\eqref{eq:residual-bin-cib}, and this term correlates with $I$ through the power of
$I$ itself. The measured cross-spectrum is therefore
\begin{equation}
    C_\ell^{g_{\rm raw} I} = C_\ell^{\delta_g I} + A_{\rm ELG}\, C_\ell^{II} ,
    \label{eq:craw-I}
\end{equation}
and we obtain the first term by subtracting the second, which we know from the
measured response and the measured $C_\ell^{II}$. Here we assume that the
Galactic part of the map error does not correlate with $I$. Note that $C_\ell^{II}$ and $C_\ell^{\kappa I}$ are properties of the map and carry over directly to any sample. In contrast, $C_\ell^{\delta_g I}$ depends on the bias and the redshift of the sample, and we use the ELG value for every bin with the imprint.

Our assumption that the correction leaves $A_i\, I$ untouched needs a comment,
since a correction can only remove what its template contains. For a template
without extragalactic signal the assumption is exact. The absolute maps SFD98
and GNILC do contain the signal, but their fitted slope is much smaller than
$A$, so they remove only a small part of $I$ (Appendix~\ref{app:derivation}). A
reddening-difference template built from SFD98 contains $I$ and removes more of it (Eq.~\ref{eq:app-ckg-diff}). For the entry that represents the DESI weights the three terms above are therefore upper bounds.

The measured band powers are interpolated with a smooth curve and held constant beyond the first and last band centers. None of these spectra is measured below $\ell = 30$. This matters only for the forecast for $f_{\rm NL}$, which uses $C_\ell^{\kappa g}$ down to $\ell = 8$. There we take $C_\ell^{\kappa I}$ to be proportional to the model $C_\ell^{\kappa g}$ of the ELG sample, with one amplitude fitted to the three measured bands. A single amplitude does not describe the three bands perfectly ($\chi^2 = 9.1$ for two degrees of freedom), so this extension is approximate. These multipoles carry $3$--$4\%$ of the shift of $f_{\rm NL}$ for the full data vector.

\paragraph{Parameter bias.}
A term $\Delta\mathbf{C}_\ell$ in the data vector that is not in the model shifts the best-fit parameters by \cite{amara-refregier-fisher-2008}:

\begin{equation}
    \Delta\theta_a = \sum_b \left(F^{-1}\right)_{ab}
    \sum_\ell \frac{\partial \mathbf{C}_\ell^{\rm T}}{\partial \theta_b}\,
    \mathbf{Cov}_\ell^{-1}\, \Delta\mathbf{C}_\ell ,
    \label{eq:fisher-bias}
\end{equation}
where $\mathbf{C}_\ell$ collects the spectra of
Table~\ref{tab:forecast-vectors}, $\mathbf{Cov}_\ell$ is their covariance and
$F$ is the Fisher matrix of the parameters $\boldsymbol{\theta}$ including the
priors. We quote the shift in units of the marginalized uncertainty $\sigma$ of
each parameter. The shift is linear in $\Delta\mathbf{C}_\ell$, so it separates
exactly into the contributions of the Galactic term, the CIB term in
$C_\ell^{\kappa g}$ and the CIB terms in $C_\ell^{gg}$.

We also ask how small the response of a sample has to be for the shift to stay
below a threshold, and we call this value $A_{\rm crit}$. We use two thresholds --- $0.5\sigma$ and $1\sigma$. For the Galactic term, the shift scales as
$A^2$, so $A_{\rm crit} = A_{\rm ELG}\, [t\,\sigma / |\Delta\theta|_{A_{\rm ELG}}]^{1/2}$
for a threshold of $t\,\sigma$. For the Galactic and the CIB terms, the CIB terms are linear and
quadratic in $A$, and we solve for $A_{\rm crit}$ numerically.

\paragraph{Effects left out.}
Two effects are left out. The galaxy bias is linear, although the clustering of
emission-line galaxies is known to be more complex \cite{karim2026-elg-dr3};
Section~\ref{sec:results-fnl} tests one extension. We also do not model the
removal of real large-scale power by the regression itself, through chance
correlations between the templates and the galaxy field. In $f_{\rm NL}$
analyses this is calibrated with mocks
\cite{mehdi2024-lrg, alberto-2024-elg-sysnet}. It acts on the same multipoles as
the imprint with the opposite sign, and it depends on the templates used, so it
does not cancel between corrections. The regression also removes real power
through a correlation that is not chance, between the galaxies and the
extragalactic signal in templates that contain it. Standard mocks do not
include this effect, and we do not model it beyond the bound on the CIB terms
above.

\section{Results: the impact of reddening errors on inflation, standard cosmology and the neutrino mass}
\label{sec:results-forecast}

\subsection{Bias on $f_{\rm NL}$}
\label{sec:results-fnl}
\begin{figure}[h]
    \centering
    \includegraphics[width=\linewidth]{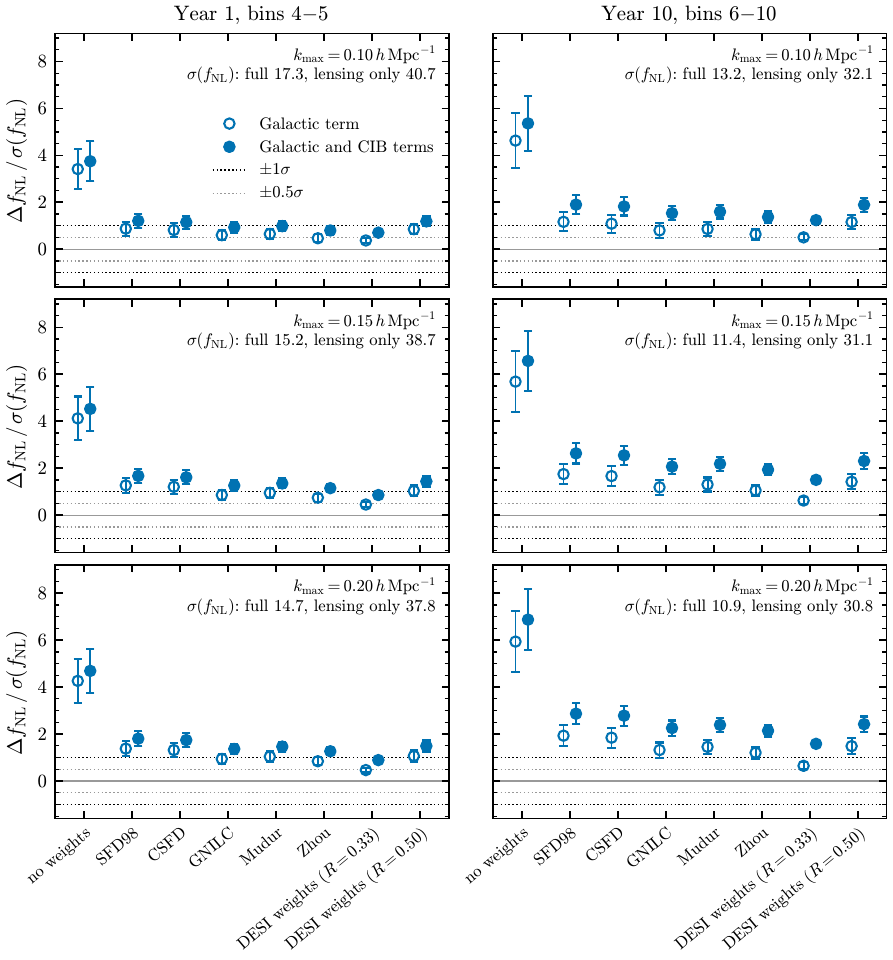}
    \caption{Bias on $f_{\rm NL}$, in units of its forecast uncertainty, from the imprint left by each correction (horizontal axis), for a sample with the ELG response in a Rubin LSST $\times$ Simons Observatory analysis. The columns show the Year~1 and Year~10 lens bins and the rows show the three
    scale cuts. Open markers show the bias from the Galactic term alone and filled markers the bias from the Galactic and CIB terms together. The two entries for the DESI weights use the two ends of the measured range of $R$ (Section~\ref{sec:template-limits}). Black and gray dotted lines mark $\pm 1\sigma$ and $\pm 0.5\sigma$. Each panel gives $\sigma(f_{\rm NL})$ for the full data vector and for the cross-spectra with lensing alone. Error bars propagate the jackknife uncertainty of the measured leftover power. They are strongly correlated between entries.}
    \label{fig:fnl-forecast}
\end{figure}

First, we look at the impact of the imprint on the amplitude of primordial
non-Gaussianity, $f_{\rm NL}$, in Figure~\ref{fig:fnl-forecast}. It shows the
bias from the imprint left by each correction, in units of the forecast
uncertainty. For $k_{\rm max} = 0.20\ h\,{\rm Mpc}^{-1}$ the uncertainty is
$\sigma(f_{\rm NL}) = 14.7$ in Year~1 and $10.9$ in Year~10 for the full data
vector. These values are larger than the current bound from the CMB
\cite{planck2020-fnl-best} because our setup is conservative. We use the galaxy
auto-spectra only from $\ell = 30$ and only the high-redshift bins.

Each result depends on two choices, the map used to select the sample and the
correction applied afterwards. Our sample is always selected with SFD98, and
the entries of Figure~\ref{fig:fnl-forecast} differ in the correction. The five
corrected entries use the same regression on imaging templates and differ only
in the reddening map that is added to it. We refer to them by that map, e.g.,
corrected with SFD98. The entry with no weights is the raw density of the
sample, and the two entries for the DESI weights rescale its leftover power by
$R^2$ (Eq.~\ref{eq:dCgg-R}). As a reminder, this means that the last two
entries represent a correction with reddening-difference templates, while the
five corrected entries before them use absolute reddening maps.

The bias of each entry is the sum of two contributions
(Section~\ref{sec:imprint-injection}). The Galactic term is the bias from the
leftover power in the auto-spectrum. The CIB terms are the bias from the
extragalactic signal in SFD98. Since our sample is selected with SFD98, both
are present, and the filled markers show their sum. The open markers show the
Galactic term alone. This is the bias we expect for a sample selected with a
map that has the Galactic error of SFD98 but not its extragalactic signal, such
as CSFD. We do not have such a sample, and this reading assumes that CSFD and
SFD98 share their Galactic error, which follows from how CSFD is built.

\paragraph{Size of the bias.}
For the sample corrected with SFD98, which is the common practice, the Galactic term biases $f_{\rm NL}$ by $1.4\sigma$ in Year~1 and $1.9\sigma$ in Year~10. The CIB terms add $0.4\sigma$ and $0.9\sigma$. They add the same amount to every entry, since they do not depend on the correction, which is why the filled markers sit at a constant offset above the open ones.

The other absolute reddening maps do not change this picture much. The sample
corrected with CSFD has the same Galactic term as with SFD98. With GNILC and
with Mudur it is somewhat lower. Note that the entries share the same galaxy map, so their uncertainties are
strongly correlated. The error bars in Figure~\ref{fig:fnl-forecast} are
correct for each entry on its own, but they overstate the uncertainty of a
difference between two entries. When we compute the uncertainty of the
difference directly, the Galactic term with GNILC is lower than with SFD98 by
$0.44 \pm 0.13\sigma$ in Year~1, and with Mudur by $0.34 \pm 0.16\sigma$. These
two maps therefore do leave a smaller imprint than SFD98, although the
individual error bars alone would not show it.

We note additional caveats for two of the cases. First, the case with no weights is also not corrected for depth, seeing or stellar density, so part of its bias may come from these and not from the reddening error. Second, the case corrected with the Zhou map is likely too low. We measure the leftover power along a template that contains the Zhou map itself, and a regression removes most of what lies along its own map. With a template that does not contain it, this entry rises from $0.84 \pm 0.18\sigma$
to $1.03 \pm 0.45\sigma$ in Year~1 (Section~\ref{sec:regression-abs-ebv-residual}). Its difference from the sample corrected with SFD98 is then within the uncertainty.

\paragraph{Size of the systematic shift in absolute $f_{\rm NL}$.}
It is instructive to look at the systematic shift in units of $f_{\rm NL}$. For the sample corrected with SFD98 the Galactic term is $\Delta f_{\rm NL} \approx 20$ in Year~1 and $21$ in Year~10. The two values are nearly the same although Year~10 has more than twice the number density and a larger area. The imprint is an additive term in the auto-spectrum, so more data do not average it down. This is why the shift in units of $\sigma$ grows from Year~1 to Year~10. The uncertainty shrinks and the imprint does not. For comparison, a shift of $20$ is four times the uncertainty of the current CMB bound.

We also find that the shift of $f_{\rm NL}$ is always positive. A positive
$f_{\rm NL}$ raises the galaxy bias of a sample with $b > 1$ on the largest
scales, and therefore adds power at low multipoles. The leftover power of the
imprint also makes up a larger share of the clustering at low multipoles than
at high ones. A model without the imprint therefore reads it as a positive
$f_{\rm NL}$. For the same reason the shift of $f_{\rm NL}$ follows the
leftover power at low multipoles and not its integral over all scales. It grows
with the scale cut, from about $0.9\sigma$ at
$k_{\rm max} = 0.10\ h\,{\rm Mpc}^{-1}$ to $1.4\sigma$ at $0.20$ in Year~1, and
changes little above $0.15$.

Most of the systematic shift comes from the Galactic term. For the sample corrected with
SFD98, the leftover power in the auto-spectrum carries $76\%$ of the total shift
in Year~1 and $67\%$ in Year~10. The CIB term in $C_\ell^{\kappa g}$ contributes
$15\%$ and $26\%$, and the CIB terms in the auto-spectrum contribute the rest.

\paragraph{$f_{\rm NL}$ from only $C^{\kappa g}_{\ell}$.}
If only the cross-spectra with lensing are used, the Galactic term does not contribute to the systematic shift at all. The shift on $f_{\rm NL}$ from the CIB terms is below $0.3\sigma$
in both years. The CIB term is instead absorbed by the linear galaxy bias of each bin,
which shifts by about $2\sigma$. This protection comes at a cost, since
$\sigma(f_{\rm NL})$ grows to $37.8$ in Year~1 and $30.8$ in Year~10, a factor
of $2.6$ to $2.8$ larger than for the full data vector.

In units of $f_{\rm NL}$ the picture is less comfortable. The multipoles at
$\ell \ge 30$, where $C_\ell^{\kappa I}$ is measured, shift $f_{\rm NL}$ by
about $-8$ in Year~1 and $-9$ in Year~10. This is a third of the systematic shift of the
full data vector, and it only looks small in units of $\sigma$ because the
uncertainty is large. Below $\ell = 30$ we do not measure $C_\ell^{\kappa I}$
and assume that it follows the shape of $C_\ell^{\kappa g}$
(Section~\ref{sec:imprint-injection}). With this assumption these multipoles
shift $f_{\rm NL}$ by about $+6$ and $+7$, and the net shift is $-2$ in both
years. The near cancellation therefore rests on the extrapolated part, and most
of the information on $f_{\rm NL}$ in the cross-spectrum lies at those
multipoles. A cross-spectrum analysis with a lower uncertainty than ours would
need to measure $C_\ell^{\kappa I}$ there.

\begin{table}[h]
\centering
\small
\begin{tabular}{lcc}
\hline
Configuration & $\sigma(f_{\rm NL})$ & $\Delta f_{\rm NL}/\sigma$ \\
 & Year 1 / Year 10 & Year 1 / Year 10 \\
\hline
Full data vector (headline) & $14.7$ / $10.9$ & $1.38 \pm 0.31$ / $1.93 \pm 0.44$ \\
Limber, no RSD, all spectra from $\ell = 30$ & $14.5$ / $10.2$ & $1.57 \pm 0.36$ / $2.28 \pm 0.52$ \\
Tighter priors on $n(z)$ \cite{fang2021-lsst-so-forecast-cmb} & $14.3$ / $10.8$ & $1.45 \pm 0.32$ / $2.04 \pm 0.45$ \\
Leftover power measured with two stellar maps & $14.7$ / $10.9$ & $1.30 \pm 0.57$ / $1.87 \pm 0.79$ \\
Free $k^2$ bias term, Eq.~\eqref{eq:bias-k2} & $18.0$ / $13.7$ & $0.77 \pm 0.30$ / $1.04 \pm 0.41$ \\
\hline
\end{tabular}
\caption{Dependence of the $f_{\rm NL}$ result on the modeling. The bias is the
Galactic term for the sample corrected with SFD98, at
$k_{\rm max} = 0.20\ h\,{\rm Mpc}^{-1}$.}
\label{tab:fnl-robustness}
\end{table}

\paragraph{Dependence on the modeling.}
Table~\ref{tab:fnl-robustness} shows how the result depends on the modeling.
The simpler calculation in the second row, which uses the Limber approximation
without redshift-space distortions and starts all spectra at $\ell = 30$, gives
a systematic bias that is higher by $0.2$--$0.35\sigma$. On the same multipoles the Limber
approximation alone lowers $\sigma(f_{\rm NL})$ by $5$--$7\%$ and changes the
systematic bias by about $0.1\sigma$. Tighter priors on the redshift distributions change
$\sigma(f_{\rm NL})$ by $3\%$ or less and the systematic shift by about $0.1\sigma$. A
second estimate of the leftover power, which uses two stellar maps, gives a consistent shift with about twice the uncertainty. Magnification changes the systematic shift by at most $0.03\sigma$.

The largest change comes from the model of the galaxy bias. A free $k^2$ term
in the galaxy bias of each bin reduces the Galactic term in the shift of
$f_{\rm NL}$ to $0.77\sigma$ in Year~1 and $1.04\sigma$ in Year~10, but it does so
at a cost. The uncertainty on $f_{\rm NL}$ grows by $22\%$ and $25\%$, and part
of the reduction in units of $\sigma$ is due to this larger uncertainty. In
absolute terms the shift falls from $\Delta f_{\rm NL} \approx 20$ to $14$ in
Year~1 and from $21$ to $14$ in Year~10. The new parameters shift by about half
of their own uncertainty, so the imprint that they absorb would not be noticed
in them. A more flexible model of the galaxy bias therefore reduces the shift
of $f_{\rm NL}$ but does not remove it.

\subsection{Bias on $\sigma_8$, $\Omega_m$ and $\Sigma m_\nu$}
\label{sec:results-amplitude}

\begin{figure}[ht]
    \centering
    \includegraphics[width=\linewidth]{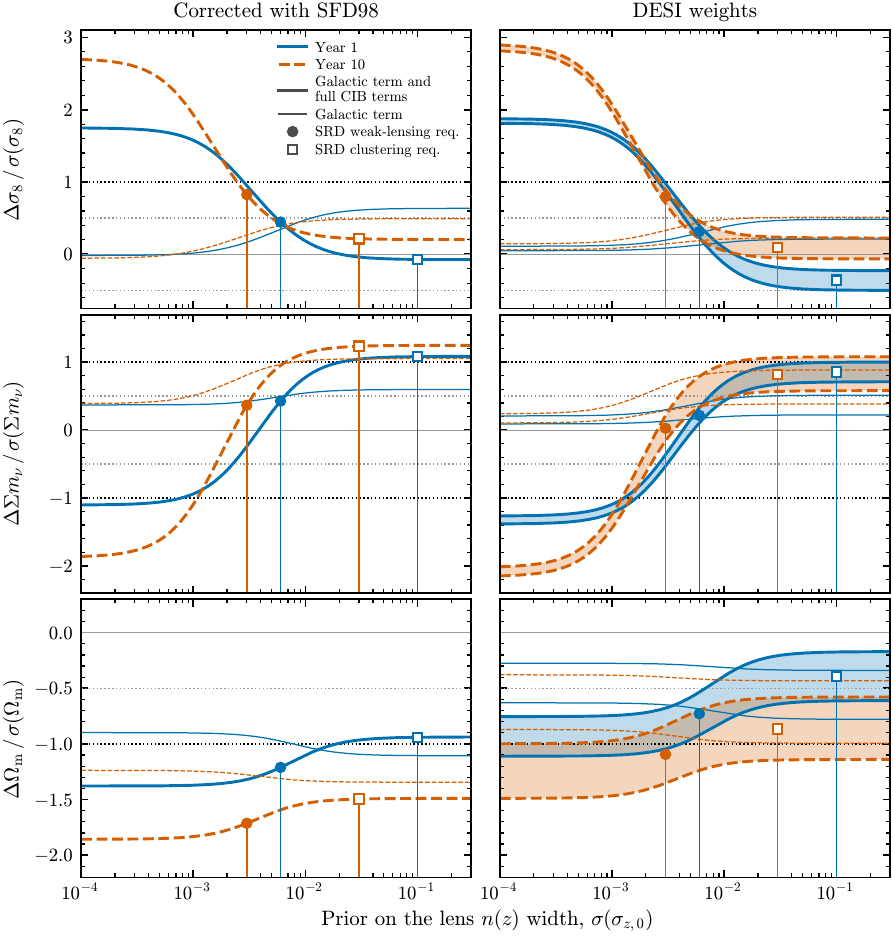}
    \caption{Shifts of $\sigma_8$ (top), $\Sigma m_\nu$ (middle) and $\Omega_m$
    (bottom), in units of their uncertainties, as a function of the prior on
    the width $\sigma_{z,0}$ of the lens redshift distributions, for a sample
    with the ELG response that is selected with SFD98. The left column is for a
    correction with SFD98 and the right column for the DESI weights, with bands
    for $R = 0.33$ to $0.50$. Thin lines show the Galactic term alone and thick
    lines the Galactic term with the full CIB terms. For the DESI weights the
    result lies between the two. Circles and squares mark the SRD requirements
    for the weak-lensing and clustering analyses, and dotted lines mark
    $0.5\sigma$ and $1\sigma$.}
    \label{fig:nz-prior}
\end{figure}

We now turn to the parameters of the standard model and the neutrino mass. The
setup differs from that for $f_{\rm NL}$ in that all lens bins enter, with the
imprint in the high-redshift bins only (Section~\ref{sec:forecast-setup}).

The shifts of $\sigma_8$ and $\Sigma m_\nu$ depend on how well the redshift
distributions of the lens bins are known. We therefore show them in
Figure~\ref{fig:nz-prior} as a function of the prior on the width of the
redshift distributions, and not as single numbers. We describe the photometric
redshift scatter of the lens sample as $\sigma_{z,0}(1+z)$ with a fiducial
$\sigma_{z,0} = 0.03$ and one value of $\sigma_{z,0}$ for all bins
(Section~\ref{sec:forecast-vectors}).

We show two corrections. The left column of Figure~\ref{fig:nz-prior} is for
the sample corrected with SFD98, which is the common practice and stands for
the absolute reddening maps. The right column is for the DESI weights, which
stand for the reddening-difference templates. We first describe the sample
corrected with SFD98, for which Table~\ref{tab:amplitude-terms} gives the
contribution of each term, and then turn to the DESI weights.

\begin{table}[h]
\centering
\small
\begin{tabular}{llcccc}
\hline
 & Prior on $\sigma_{z,0}$ & Galactic & CIB in $C_\ell^{\kappa g}$ & CIB in $C_\ell^{gg}$ & Total \\
\hline
$\Delta\sigma_8/\sigma$ & fixed & $-0.02$ / $-0.06$ & $+1.97$ / $+3.03$ & $-0.20$ / $-0.27$ & $+1.75$ / $+2.71$ \\
 & $0.006$ / $0.003$ & $+0.35$ / $+0.27$ & $+0.05$ / $+0.55$ & $+0.05$ / $+0.01$ & $+0.44$ / $+0.82$ \\
 & none & $+0.64$ / $+0.49$ & $-0.90$ / $-0.43$ & $+0.19$ / $+0.14$ & $-0.08$ / $+0.20$ \\
$\Delta\Omega_m/\sigma$ & fixed & $-0.90$ / $-1.24$ & $-0.42$ / $-0.53$ & $-0.06$ / $-0.09$ & $-1.38$ / $-1.86$ \\
 & $0.006$ / $0.003$ & $-0.97$ / $-1.28$ & $-0.13$ / $-0.31$ & $-0.11$ / $-0.12$ & $-1.21$ / $-1.71$ \\
 & none & $-1.11$ / $-1.34$ & $+0.35$ / $+0.03$ & $-0.18$ / $-0.17$ & $-0.94$ / $-1.49$ \\
$\Delta\Sigma m_\nu/\sigma$ & fixed & $+0.37$ / $+0.39$ & $-1.70$ / $-2.56$ & $+0.23$ / $+0.30$ & $-1.11$ / $-1.87$ \\
 & $0.006$ / $0.003$ & $+0.50$ / $+0.81$ & $-0.16$ / $-0.63$ & $+0.09$ / $+0.18$ & $+0.43$ / $+0.37$ \\
 & none & $+0.60$ / $+1.05$ & $+0.45$ / $+0.04$ & $+0.04$ / $+0.15$ & $+1.09$ / $+1.25$ \\
\hline
\end{tabular}
\caption{Contributions to the shifts of $\sigma_8$, $\Omega_m$ and
$\Sigma m_\nu$ for the sample corrected with SFD98, for Year~1 / Year~10, at
three priors on the width of the lens redshift distributions. The middle row
of each block is the SRD requirement for the source sample. The values for
$\sigma_8$ and $\Omega_m$ are from the fit without neutrino mass and those for
$\Sigma m_\nu$ from the fit that includes it.}
\label{tab:amplitude-terms}
\end{table}

\paragraph{When the redshift distributions are well known.}
The CIB term adds power to $C_\ell^{\kappa g}$ that no galaxy in the bin
produces. With the width of the redshift distributions fixed, the fit can only
account for it with a higher amplitude of matter fluctuations. The shift of
$\sigma_8$ is then $+1.8\sigma$ in Year~1 and $+2.7\sigma$ in Year~10. It comes
from the CIB term in $C_\ell^{\kappa g}$, while the Galactic term alone moves
$\sigma_8$ by less than $0.1\sigma$ (Table~\ref{tab:amplitude-terms}).

The neutrino mass responds in the opposite direction. The primordial amplitude
is fixed by the CMB prior, so a higher amplitude today is read as less
suppression of growth, i.e., a lower neutrino mass. $\Sigma m_\nu$ shifts by
$-0.11$~eV in Year~1 and $-0.17$~eV in Year~10. For our fiducial value of
$0.06$~eV the recovered mass is formally negative.

\paragraph{When the redshift distributions are poorly known.}
If the width is free, the fit can also account for the extra power in
$C_\ell^{\kappa g}$ by changing the redshift distributions. The CIB term in
$C_\ell^{\kappa g}$ then no longer raises $\sigma_8$. It changes sign, and it
partly cancels the Galactic term, which now shifts $\sigma_8$ by about
$+0.5\sigma$. The net shift of $\sigma_8$ is below $0.3\sigma$.

The neutrino mass is now biased high, by about $+0.25$~eV in both years. Most of this comes from the Galactic term, which pushes the neutrino mass up for any prior. Note that the positive shift of $\sigma_8$ from the Galactic term in the previous paragraph is from the fit without neutrino mass. There $\Omega_m$ is free apart from the prior on $h$, and the leftover power mostly lowers $\Omega_m$ (Table~\ref{tab:amplitude-terms}). In the fit with neutrino mass, the CMB priors on $\ln A_s$ and $\omega_c$, together with the prior on $h$, tightly constrain $\Omega_m$, so the leftover power acts on the amplitude instead. It raises $C_\ell^{gg}$ of the high-redshift bins and leaves $C_\ell^{\kappa g}$ unchanged, which lowers the amplitude inferred from the two together (Eq.~\eqref{eq:amp-ratio}). A lower amplitude is read as more suppression of growth. The Galactic term therefore always raises the neutrino mass, and the CIB term in $C_\ell^{\kappa g}$ decides the sign of the total. The imprint can thus bias the neutrino mass in either direction.

\paragraph{The transition.}
The shift of $\sigma_8$ reaches $0.5\sigma$ when the prior on $\sigma_{z,0}$ is
about $0.005$ and $1\sigma$ when it is about $0.003$. Since the fiducial value
is $\sigma_{z,0} = 0.03$, this corresponds to knowing the photometric redshift
scatter of the lens sample to about $17\%$ and $10\%$ of its value. The shift
of $\Sigma m_\nu$ changes sign at $0.004$ in Year~1 and $0.002$ in Year~10. The
SRD requirements for the source sample, which
\cite{fang2021-lsst-so-forecast-cmb} adopt for the lens sample, are $0.006$ and
$0.003$ and lie in this range. At these values the shift of $\sigma_8$ is
$0.4\sigma$ and $0.8\sigma$ and that of $\Sigma m_\nu$ is $0.4\sigma$ in both
years. We caution that the shifts are small there because the CIB term in
$C_\ell^{\kappa g}$ passes through zero in this range, and that the position of
this zero depends on the parameterization. With one width per bin and the same
prior on each, the shift of $\sigma_8$ at these values is $0.9\sigma$ and
$1.8\sigma$.

\paragraph{The matter density.}
The shift of $\Omega_m$ behaves differently, and more like that of
$f_{\rm NL}$. For the sample corrected with SFD98 it is between $-0.9\sigma$
and $-1.9\sigma$ for any prior on $\sigma_{z,0}$, as the nearly flat curves in
the bottom row of Figure~\ref{fig:nz-prior} show. Most of it comes from the
Galactic term (Table~\ref{tab:amplitude-terms}). The leftover power of the
imprint is largest at low multipoles, so it changes the shape of the
auto-spectrum, and $\Omega_m$ is the parameter that sets this shape in our fit.
Note that we hold $n_s$ and $\omega_b$ fixed and use a prior on $h$
(Table~\ref{tab:forecast-vectors}), so $\Omega_m$ is the only shape parameter
that can respond. In an analysis with more freedom in the shape, or with a CMB
prior on $\Omega_m$, this shift would be shared with other parameters or
reduced.

\paragraph{Does a better correction help?}

Here we explore the impact of using the DESI-style reddening-difference template method and we find that the answer is different for the Galactic term and the CIB term.

We find that the Galactic term becomes smaller. Among the absolute reddening maps, the samples corrected with SFD98 and with CSFD have the largest Galactic term in $\Omega_m$, between $-0.9\sigma$ and $-1.3\sigma$. With GNILC, Mudur or Zhou it is between $-0.6\sigma$ and $-1.0\sigma$ (Table~\ref{tab:amplitude-corrections}). The entries share the same galaxy map, so their uncertainties are strongly correlated. Taking this into account, the samples corrected with GNILC and with Mudur have a Galactic term in $\Omega_m$ that is smaller than with SFD98 by $0.2$--$0.4\sigma$, at about three times the uncertainty of the difference. For the sample corrected with the Zhou map the comparison is not conclusive, since its unbiased estimate has a larger uncertainty. None of the absolute maps brings the Galactic term in $\Omega_m$ below $0.5\sigma$. The DESI weights do better only at the lower end of the measured range of $R$, which is the optimistic end (Section~\ref{sec:imprint-injection}). With $R = 0.33$ their Galactic term in $\Omega_m$ is $-0.3\sigma$ in Year 1 and $-0.4\sigma$ in Year 10, while with $R = 0.50$ it is the same as with GNILC or Mudur. The same holds for $\sigma_8$, where the Galactic term of the DESI weights is smaller than with SFD98 only for $R = 0.33$. Over all priors on the redshift distributions, the Galactic term of the DESI weights stays at or below about $0.5\sigma$ in $\sigma_8$ and between $-0.3\sigma$ and $-1.0\sigma$ in $\Omega_m$ (right column of Figure~\ref{fig:nz-prior}, thin bands).

The CIB terms are a different matter. A correction can only remove what its
template contains. CSFD, Mudur and Zhou contain no extragalactic signal, so the
CIB terms remain in full for the samples corrected with them. SFD98 and GNILC
do contain it, and a regression on them removes a part of it. This part is
small, because the fitted slope of an absolute map is much smaller than the
response (Section~\ref{sec:imprint-injection}), and we neglect it. For these
two maps the CIB terms that we show are therefore slightly too large. With this
caveat, the shifts of $\sigma_8$ and $\Sigma m_\nu$ for well-known redshift
distributions are the same for all five maps to within $0.2\sigma$.

The templates of the DESI weights do contain the extragalactic signal and
remove part of it, by an amount that we have not measured. We therefore show
two limits in the right column of Figure~\ref{fig:nz-prior}. If the templates
removed all of the extragalactic signal, only the Galactic term would remain
(thin bands). If they removed none of it, the full CIB terms would add to it
(thick bands). The shifts are linear in the injected terms, so the result for
any fraction in between lies between the two bands. With well-known redshift
distributions the two limits are far apart. For $\sigma_8$ they are $0.1\sigma$
and $1.8\sigma$ in Year~1 and $0.1\sigma$ and $2.9\sigma$ in Year~10. For this
kind of correction the bias on $\sigma_8$ and on the neutrino mass is therefore
set by how much of the extragalactic signal is left in the sample. This is the
quantity that needs to be measured next.

\paragraph{A note on the size of the shifts.}
All shifts are computed in the linear approximation of
Eq.~\eqref{eq:fisher-bias}. This approximation does not know that the neutrino
mass cannot be negative. Where we find a negative shift of $\Sigma m_\nu$ that
is larger than the fiducial mass, a real analysis would not return a negative
mass. Its posterior would instead pile up against the boundary at zero (unless neutrino mass prior is allowed to be negative). The
values should therefore be read as the size and direction of the effect, and
not as the outcome of a fit.

\subsection{Requirements on the response and on the redshift calibration}
\label{sec:requirement}

The results above are for one response, that of the ELG sample. The shifts
scale with it, as $A^2$ for the Galactic term, as $A$ for the CIB term in
$C_\ell^{\kappa g}$, and as a mix of $A$ and $A^2$ for the CIB terms in
$C_\ell^{gg}$ (Section~\ref{sec:imprint-injection}). We can therefore ask how
small the response has to be for the bias on $f_{\rm NL}$ to stay below a
threshold. Table~\ref{tab:acrit} gives this value, $A_{\rm crit}$, for
thresholds of $0.5\sigma$ and $1\sigma$.

\begin{table}[h]
\centering
\small
\begin{tabular}{llcc}
\hline
Correction & Terms & Year 1 & Year 10 \\
\hline
SFD98 & Galactic & $5.7$ / $8.1$ & $4.8$ / $6.8$ \\
 & Galactic and CIB & $4.4$ / $6.8$ & $3.1$ / $4.9$ \\
DESI weights, $R = 0.50$ & Galactic & $6.5$ / $9.2$ & $5.5$ / $7.8$ \\
 & Galactic and CIB & $4.9$ / $7.5$ & $3.3$ / $5.4$ \\
DESI weights, $R = 0.33$ & Galactic & $9.8$ / $13.9$ & $8.3$ / $11.8$ \\
 & Galactic and CIB & $6.4$ / $10.2$ & $4.0$ / $6.8$ \\
\hline
\end{tabular}
\caption{Largest response $A_{\rm crit}$, in units of per mag, for which the
bias on $f_{\rm NL}$ stays below $0.5\sigma$ / $1\sigma$, at
$k_{\rm max} = 0.20\ h\,{\rm Mpc}^{-1}$. The ELG sample has $A = 9.48$ per mag.
For the DESI weights the rows with the CIB terms use the full CIB terms and are
therefore lower limits on $A_{\rm crit}$.}
\label{tab:acrit}
\end{table}

For a sample corrected with SFD98 the response has to be below about $5$ per
mag for a bias under $0.5\sigma$, and below about $3$ per mag in Year~10 if the
sample is also selected with SFD98. The ELG response of $9.48$ per mag exceeds
these limits by a factor of $1.7$ to $3.1$. Even for the DESI weights at the
favorable end of the measured range of $R$, the ELG response is only allowed in
Year~1 and only if the CIB terms are absent.

A sample with $A = 1.4$ per mag, the value we estimate for a magnitude limit in
the $i$ band, has a much smaller bias on $f_{\rm NL}$. For a sample corrected
with SFD98 the Galactic term is below $0.05\sigma$ in both years. If the sample
is selected with SFD98, the CIB terms bring the total to $0.09\sigma$ in Year~1
and $0.18\sigma$ in Year~10. With no weights the total is $0.16\sigma$ and
$0.27\sigma$. The LSST gold sample as a whole therefore passes. Note that at
this response most of the bias comes from the CIB terms, which are linear in
the response, and not from the Galactic term.

Two caveats apply to the gold sample. A tomographic bin is not the whole
sample. The edges of a photometric-redshift bin are boundaries in color, and an
error of the reddening moves galaxies across them. The responses of the bins
can be larger than that of the sample, and of either sign, and only their
weighted sum is fixed. Samples built to reach high redshift, such as
Lyman-break galaxies, are selected near faint limits in the bluest bands, where
the extinction coefficients are largest. Their response may be as large as that
of the ELG sample or larger, but we have not measured it.

The second requirement is on the redshift calibration. For $\sigma_8$, Figure~\ref{fig:nz-prior} shows that the imprint matters once
the photometric redshift scatter of the lens sample is known to better than
about $10\%$ of its value, which is the level of the SRD requirement for Year~10. A better
redshift calibration therefore makes an analysis more sensitive to the imprint
and not less.

The response of any sample can be measured as in Section~\ref{sec:response},
by shifting the reddening by a constant, repeating the selection including the
photometric redshifts and the binning, and counting. We recommend this test for
every tomographic bin of a clustering analysis, with $A_{\rm crit}$ as the
number to compare with.

\section{Discussion and Conclusion}
\label{sec:discussion-conclusion}

The cosmology community has been looking forward to both the Vera C. Rubin Observatory and the Simons Observatory data for the Stage-IV cross-correlation analysis for over a decade now, and we expect the first results of these collaborations in the next few years. It is therefore of great importance that our modeling accuracy goes hand in hand with the precision we will be gaining. While our forecast has been restricted to LSST and SO, the issue of the selection function affects any ongoing or future galaxy survey that selects its samples on dereddened photometry, to a degree set by the response of each sample. In the DESI photometric ELG sample, the choice of reddening map already changes the tension with \textit{Planck} in $\sigma_8$--$\Omega_m$ by $1.2\sigma$~\cite{karim2025-elg-cmb-lensing}, and the DESI DR1 weights leave between a third and a half of the amplitude of the Galactic imprint in the angular density of the spectroscopic ELG sample (Section~\ref{sec:template-limits}). Current analyses that rely on the large-scale clustering of samples with a large response, such as measurements of $f_{\rm NL}$ or of the shape of the power spectrum, may therefore already be affected, although we have not assessed this here. As we push to higher redshifts with the help of bluer bands, it is now the right time to revisit the old problem of Milky Way extinction.

In this paper we studied the error of the Milky Way reddening map that is used to select a galaxy sample, and how it may affect the measurement of primordial non-Gaussianity, of the growth of structure and of the neutrino mass. Because galaxies are selected on dereddened photometry, the error of the map becomes part of the selection function. A correction that is applied afterwards can only remove the part of this imprint that its template traces. We modeled the imprint to first order and measured it in the photometric emission-line galaxies of the DESI Legacy Imaging Surveys, for regressions on five reddening maps and for the DESI weights, and with the \textit{Planck} PR4 and ACT DR6 CMB lensing maps. We then propagated what is left after each correction to a Rubin LSST $\times$ SO analysis for a sample
with the same response.

\paragraph{Our key findings from data:}
\begin{itemize}
    \item The response of a sample to a reddening error is set by its selection and can be measured. Shifting the reddening in the catalog and repeating the selection gives $A = 9.48$ per mag for the DESI DR$9$ photometric ELG sample, i.e., an error of $10$~mmag in $E(B-V)$ changes its density by $9.5\%$. Most of this comes from the faint limit on the $g$-band fiber magnitude. A sample limited in a red band has a response several times smaller. In comparison, the DR1 spectroscopic sample, whose color box differs in one cut, has a response of $10.8$ per mag.
    
    \item A regression on an absolute reddening map removes only part of the Galactic imprint. With no correction, the imprint makes up $10$--$13\%$ of the measured galaxy auto-spectrum at $50 \le \ell < 400$. After a correction with SFD98, CSFD, GNILC or the Mudur map it still makes up $4$--$9\%$. Note that each correction is a regression on a set of imaging templates to which the reddening map is added, so these numbers describe the full correction and not the reddening map alone. The reason so much is left is that the error of the map does not fully follow the dust column, and an absolute map can remove only the part of the error that correlates with the dust column.
    
    \item A reddening-difference template is also not sufficient for precision cosmology. For the weights of the DESI DR1 spectroscopic ELG sample, between a third and a half of the amplitude of the Galactic imprint is left. We quantified this with two templates that differ from the ones in the official DESI regression. The first, $E_{\rm CSFD} - E_{\rm Mudur}$, uses a stellar map that DESI did not use and shows that about half of the amplitude is left. The second, $E_{\rm CSFD} - E_{\rm HI}$, shows that about a third of the amplitude is left, which is optimistic since the H\,\textsc{i} map is one of the maps that DESI fits. For a sample like the DR9 photometric one, these leftovers would correspond to 1--3\% of $C_\ell^{gg}$ at $50 \le \ell < 400$, i.e., the square of these fractions times the power left without weights (Table~\ref{tab:leftover}). This is the leftover power that we inject for the DESI weights in the forecast. Part of what the DESI templates, $E_{\rm SFD98} - E_{\rm Zhou}^{g-r}$ and $E_{\rm SFD98} - E_{\rm Zhou}^{r-z}$, trace is a property of the imaging, most likely the calibration of the photometry, and not of the dust. As the DESI target selection uses the same photometry as the stellar reddening map in these templates, we cannot tell how much of what the templates remove is the dust error of SFD98 and how much is an error of the Legacy Surveys photometry.
    
    \item The far-infrared reddening maps SFD98 and \textit{Planck} GNILC contain the large-scale structure. Both correlate with the ACT DR6 CMB lensing map at about $4\sigma$ after accounting for the number of tests. For a sample with the ELG response that is selected with SFD98, the part of this signal that CSFD removes adds 5--6\% to its cross-spectrum with CMB lensing. This is a lower bound if CSFD still contains extragalactic signal, and for CSFD the data are not conclusive in either direction.
\end{itemize}

\paragraph{What it implies for Rubin and SO:}
\begin{itemize}
    \item The inferred $f_{\rm NL}$ is biased high. After a correction with an absolute reddening map the Galactic imprint alone
shifts it by $1$--$2\sigma$, and for a sample selected with SFD98 the
extragalactic signal raises the systematic bias in $f_{\rm NL}$ to as much as $3\sigma$ in Year~10. With the DESI weights the Galactic imprint still shifts it by $0.5$--$1.5\sigma$. The shift is $\Delta f_{\rm NL} \approx 20$ in both years, so more data do not average it down.
    
    \item The shifts of $\sigma_8$ and of the neutrino mass depend on the redshift calibration of the lens tomographic bins. As long as the redshift distributions are poorly known, they are the dominant source of uncertainty on $\sigma_8$, and the fit uses this freedom to absorb the imprint. In that case $\sigma_8$ is nearly unbiased and the neutrino mass is biased high. Once the redshift distributions are well known, this freedom is gone and
    the imprint goes into the cosmological parameters. The extragalactic signal in SFD98 then raises $\sigma_8$ by $2$--$3\sigma$ and lowers the neutrino mass, formally below zero. The imprint starts to matter once the photometric redshift scatter is known to about $10\%$ of its value, which is the level of the SRD weak-lensing requirement for Year~10. A better redshift calibration therefore does not create the problem, but rather reveals it. For a sample with a large response that is selected with SFD98, the error of the reddening map becomes one of the next systematics to control once the redshift calibration is in hand.
    
    \item The inferred $\Omega_m$ is biased low mostly by $1$--$2\sigma$ by the
    Galactic imprint. This is because $\Omega_m$ behaves similarly to $f_{\rm NL}$. Both are read from the shape of the auto-spectrum. The leftover power of the imprint changes this shape, while the redshift calibration has little effect on it. Note that in our fit $\Omega_m$ is the only free parameter that sets this shape, so the size of the shift is specific to that choice.
    
    \item A bias on $f_{\rm NL}$ below $0.5\sigma$ requires a response below about $5$ per mag, and below about $3$ per mag in Year~10 for a sample selected with SFD98. The Rubin gold sample as a whole, with a response of about $1.4$ per mag, meets this. The response of its individual tomographic bins is not known.
\end{itemize}

\paragraph{Recommended steps for Stage-IV analyses.}
No correction that we tested removes the selection error fully, but our results point to a clear order of steps.

\begin{itemize}
    \item Measure the response of each tomographic bin, by shifting the
    reddening and repeating the selection including the photometric redshifts (Section~\ref{sec:requirement}). This costs little and tells whether the rest of this list is needed.
    
    \item Select the sample with a map that has no extragalactic signal. This removes the terms from the large-scale structure at their source and needs no fitted coefficient. For example, a Rubin clustering analysis could use, in place of SFD98, a reddening map that has been tested for extragalactic signal. The correlation with CMB lensing that we used here is one such test.
    
    \item Correct with reddening-difference templates and not with SFD98 or CSFD alone. In our tests the DESI weights --- which use such templates --- leave 30--50\% of the amplitude of the Galactic imprint, against 60--80\% for the absolute maps. They also give smaller shifts of $f_{\rm NL}$ and $\Omega_m$ than SFD98 and CSFD, although at the upper end of the measured range of $R$ these shifts are about the same as with GNILC and the Mudur map. Such templates need a second reddening map that has been tested to be free of extragalactic signal, and whose noise and calibration errors are small and independent of the galaxy photometry, at the scales and precision that the analysis requires.
    
    \item Validate the correction with an independent template. The template must not be one of the templates used in the correction, and it must not share noise or calibration with them. For the DESI weights this rules out the H\,\textsc{i} map, which is one of the templates in the regression. This is why we used a template built with the Mudur map.
    
    \item Measure how much of the extragalactic signal a correction leaves in the sample. For a DESI-style correction this fraction decides whether the shift of $\sigma_8$ is negligible or $2$--$3\sigma$.
\end{itemize}

Two further handles reduce the bias on $f_{\rm NL}$ at a cost. A free $k^2$
term in the galaxy bias lowers the shift by a third and raises the uncertainty
by a quarter. Using only the cross-spectra with CMB lensing removes the
Galactic term, but raises the uncertainty by a factor of $2.7$, and its
protection against the extragalactic signal rests on multipoles below
$\ell = 30$ where that signal has not been measured. Forward modeling of the selection with image simulations is already used by DESI
\cite{obiwan-desi, alberto-2024-elg-sysnet} and by DES \cite{balrog-desc}, and is the natural way to include the error of the reddening map in the model.

\paragraph{Going beyond ELGs.}
We measured the imprint in only the DESI DR$9$ photometric ELG and the DR$1$ spectroscopic ELG samples. However, the same concern applies to any sample that is selected on dereddened colors near a faint limit. This includes the luminous red galaxies and quasars of DESI and samples built to reach high redshift, such as Lyman-break galaxies. We defer the exploration of these other tracers to future papers, which will include analyses of full-shape and BAO clustering. Errors of the reddening map also enter other probes that use it, such as
supernova distances \cite{elliot2026-snia-dust}.

\paragraph{The southern sky.}
The stellar reddening maps used here stop at a declination of $-30^\circ$, while most of the overlap of Rubin and SO lies below that. The far-infrared and H\,\textsc{i} maps cover the southern sky, and so does the Total Galactic Extinction map of \textit{Gaia} DR3 \cite{delchambre-gaia-tge-2023}, which we did not use because its noise is too large for the map errors of interest here. Stellar maps from future \textit{Gaia} releases will improve on this, and their noise, their calibration and their extragalactic content will need the tests applied here. On the part of the sky where the H\,\textsc{i} map is valid, we also tested the two stellar maps --- \cite{rongpu-desi-dustmap-zhou} from DESI stellar spectra and \cite{nayantara-mudur-dust} from Pan-STARRS1 and 2MASS photometry. They agree with each other on the error of SFD98 more closely than either agrees with the H\,\textsc{i} map. We could not determine whether this is because the two stellar maps share an error or because the H\,\textsc{i} map and SFD98 do.

\paragraph{Limits of this work.}
The forecast rests on one sample and one footprint, on a linear galaxy bias, and on a response assigned to bins in which it has not been measured. The leftover power that we inject includes a part from the calibration of the Legacy Surveys photometry. The results for the DESI weights are inferred from the DR1 spectroscopic sample and not measured on a photometric sample corrected in that way. The shifts of $\sigma_8$ and $\Sigma m_\nu$ depend on how the redshift distributions are parameterized, and we did not model that the reddening error also changes the redshift distribution of a bin across the sky.
We have not assessed three-dimensional clustering. Whether the leftover
imprint, or the choice of stellar map, affects measurements such as baryon acoustic oscillations is the subject of separate work in preparation.

The error of SFD98 has physical causes, in the grain populations
\cite{dennis-lee-2025-end-of-fir} and in the extinction curve
\cite{singh-rv-schlafly-2026}, and will not average away. Understanding the errors of reddening maps well enough to model them, and not only to regress on them, will need the work of the interstellar-medium community as much as that of cosmologists.

\acknowledgments

We thank Emmanuel Schaan, Brandon Hensley, Susan Clark, Catherine Zucker, Joshua Speagle and Ren\'{e}e Hlo\v{z}ek for valuable discussions and input. TK was funded by the Dunlap Institute Postdoctoral Fellowship. The Dunlap Institute is funded through an endowment established by the David Dunlap family and the University of Toronto. This research was enabled in part by support provided by SciNet and the Digital Research Alliance of Canada, and all computations were done on the Trillium cluster. For part of the analysis, we used the Claude Fable, Claude Opus, Claude Sonnet and Claude Haiku large language models to help write, run and test scripts, prepare figures, and draft and edit text. The authors designed and directed the analysis, wrote the initial code and text on which this work builds, and checked all results and text, for which they take full responsibility.

This research used data obtained with the Dark Energy Spectroscopic Instrument (DESI). DESI construction and operations is managed by the Lawrence Berkeley National Laboratory. This material is based upon work supported by the U.S. Department of Energy, Office of Science, Office of High-Energy Physics, under Contract No. DE–AC02–05CH11231, and by the National Energy Research Scientific Computing Center, a DOE Office of Science User Facility under the same contract. Additional support for DESI was provided by the U.S. National Science Foundation (NSF), Division of Astronomical Sciences under Contract No. AST-0950945 to the NSF’s National Optical-Infrared Astronomy Research Laboratory; the Science and Technology Facilities Council of the United Kingdom; the Gordon and Betty Moore Foundation; the Heising-Simons Foundation; the French Alternative Energies and Atomic Energy Commission (CEA); the National Council of Humanities, Science and Technology of Mexico (CONAHCYT); the Ministry of Science and Innovation of Spain (MICINN), and by the DESI Member Institutions: www.desi.lbl.gov/collaborating-institutions. The DESI collaboration is honored to be permitted to conduct scientific research on I’oligam Du’ag (Kitt Peak), a mountain with particular significance to the Tohono O’odham Nation. Any opinions, findings, and conclusions or recommendations expressed in this material are those of the author(s) and do not necessarily reflect the views of the U.S. National Science Foundation, the U.S. Department of Energy, or any of the listed funding agencies.

The DESI Legacy Imaging Surveys consist of three individual and complementary projects: the Dark Energy Camera Legacy Survey (DECaLS), the Beijing-Arizona Sky Survey (BASS), and the Mayall z-band Legacy Survey (MzLS). DECaLS, BASS and MzLS together include data obtained, respectively, at the Blanco telescope, Cerro Tololo Inter-American Observatory, NSF’s NOIRLab; the Bok telescope, Steward Observatory, University of Arizona; and the Mayall telescope, Kitt Peak National Observatory, NOIRLab. NOIRLab is operated by the Association of Universities for Research in Astronomy (AURA) under a cooperative agreement with the National Science Foundation. Pipeline processing and analyses of the data were supported by NOIRLab and the Lawrence Berkeley National Laboratory (LBNL). Legacy Surveys also uses data products from the Near-Earth Object Wide-field Infrared Survey Explorer (NEOWISE), a project of the Jet Propulsion Laboratory/California Institute of Technology, funded by the National Aeronautics and Space Administration. Legacy Surveys was supported by: the Director, Office of Science, Office of High Energy Physics of the U.S. Department of Energy; the National Energy Research Scientific Computing Center, a DOE Office of Science User Facility; the U.S. National Science Foundation, Division of Astronomical Sciences; the National Astronomical Observatories of China, the Chinese Academy of Sciences and the Chinese National Natural Science Foundation. LBNL is managed by the Regents of the University of California under contract to the U.S. Department of Energy. The complete acknowledgments can be found at https://www.legacysurvey.org/acknowledgment/.

Based on observations obtained with \textit{Planck} (\url{http://www.esa.int/Planck}), an ESA science mission with instruments and contributions directly funded by ESA Member States, NASA, and Canada.
\appendix

\section{The response of the ELG sample}
\label{app:response}

This appendix discusses the formalism of the response $A$ of
Section~\ref{sec:imprint-model} in more detail. We first show how the response follows from the selection, then list the cuts of the two ELG samples, describe how we measure the response, and give the contribution of each cut.

\subsection{From the selection to the response}

Let $\mathcal{P}(\mathbf{m}_d)$ be the probability that an object with
dereddened magnitudes $\mathbf{m}_d$ passes the selection, and
$\phi(\mathbf{m})$ the distribution of true, dust-free magnitudes. If the map
overestimates the reddening by $s$, the dereddened magnitude in band $b$ is
brighter by $R_b\,s$, so the selected density is
\begin{equation}
    \bar{n}(s) = \int d\mathbf{m}\; \phi(\mathbf{m})\,
    \mathcal{P}\!\left(\mathbf{m} - \mathbf{R}\,s\right) .
    \label{eq:nbar-s}
\end{equation}
The response of Eq.~\eqref{eq:F} is the logarithmic derivative of this density
at $s = 0$. The error $s$ enters only through the brightening of each band, so
the derivative is a sum over bands,
\begin{equation}
    A = \left. \frac{d \ln \bar{n}}{d s} \right|_{s=0}
    = \sum_b R_b\, S_b ,
    \qquad
    S_b \equiv - \frac{1}{\bar{n}} \int d\mathbf{m}\; \phi(\mathbf{m})\,
    \frac{\partial \mathcal{P}}{\partial m_b} ,
    \label{eq:A-def}
\end{equation}
which is Eq.~\eqref{eq:A-bands}. Here $S_b$ is the fractional change of the
selected density when band $b$ alone is made brighter by one magnitude. For a
sample that is defined only by a faint limit in band $b$, the selected density
is the cumulative number count $N$ at the limit. Brightening the band by
$R_b\,s$ is then the same as moving the limit fainter by $R_b\,s$, and
\begin{equation}
    A = R_b\, \frac{d \ln N}{d m_b} ,
    \label{eq:A-maglim}
\end{equation}
i.e., the extinction coefficient times the logarithmic slope of the number
counts at the limit.

\subsection{Selection cuts of the two samples}
\label{app:color-cuts}

\begin{table}[h]
\centering
\small
\begin{tabular}{p{4.2cm} c c p{3.8cm}}
\hline
Cut & DR1 spectroscopic & DR9 photometric & Purpose \\
\hline
$g > 20$ & yes & yes & bright limit \\
$g_{\rm fib} < 24.1$ & yes & yes & faint limit, enough flux in the DESI fiber \\
$r - z > 0.15$ & yes & yes & color box \\
$g - r < 0.5\,(r-z) + 0.1$ & yes & yes & removes stars and low-redshift galaxies \\
$g - r < -1.2\,(r-z) + 1.3$ & yes & no & redshift and [O\,\textsc{ii}] selection \\
$g - r < -1.2\,(r-z) + 1.6$ & no & yes & same, wider box \\
$g_{\rm fib} > 22.6$ & no & yes & removes quasars \\
$r \leq 27$, $z \leq 25$, $r_{\rm fib} \leq 27.5$, $z_{\rm fib} \leq 25.5$, $-0.4 < g-r < 0.5$, $0.1 < r-z < 1.25$ & no & yes & well-sampled region of the training set \\
half-light radius $< 1.5''$ & no & yes & removes rare extended objects \\
Random Forest classifier & no & yes & removes interlopers and poor targets \\
spectroscopic redshift & yes & no & removes quasars and failed redshifts \\
\hline
\end{tabular}
\caption{Selection cuts of the two ELG samples. All magnitudes and colors are dereddened with SFD98. The subscript ``fib'' denotes the magnitude of the flux predicted within a fiber of $1.5''$ diameter for $1''$ Gaussian seeing. The first five rows are the DESI ELG\_LOP target selection \cite{raichoor-target-selection-elg-2023}, and the added cuts of the photometric sample are those of \cite{karim2025-elg-cmb-lensing}.}
\label{tab:color-cuts}
\end{table}

Table~\ref{tab:color-cuts} lists the cuts of the two samples. We refer to the
cuts of the photometric sample other than the classifier as the parent cuts.

\subsection{Measuring the response by re-selection}
\label{app:A-shift-estimator}

When the reddening is shifted by $\delta$, an object can leave or enter the sample, and it can do so either by crossing a parent cut or, for the
photometric sample, by changing class. For context, the DR$9$ photometric ELG catalog went through a Random Forest classifier after the parent cuts to improve purity. The three classes are true ELGs, low-redshift ($z < 0.6$) interlopers, and bad targets, and only the objects classified as true ELGs are
kept. A shift of the reddening changes the colors that the classifier sees, so it can move an object from one class to another. Our photometric catalog contains every object that passes the parent cuts with SFD98, whatever its class. Objects that leave by either route are in the catalog, and we call their number $N_{\rm leave}(\delta) = N_{\rm leave}^{\rm parent}(\delta) + N_{\rm leave}^{\rm class}(\delta)$. Objects that enter by changing class are also in the catalog, and we call their number $N_{\rm enter}^{\rm class}(\delta)$. The fourth case, objects that would enter across a parent cut, cannot be counted because these objects failed the cuts with SFD98. For these we use the objects on the other side of the same cut. The objects just outside a cut that would enter under a shift $\delta$ are about as numerous as the objects just inside it that leave under the opposite shift, $N_{\rm leave}^{\rm parent}(-\delta)$. With $N$ the number of selected objects, this gives
\begin{equation}
    A = \frac{N_{\rm enter}^{\rm class}(\delta)
              + N_{\rm leave}^{\rm parent}(-\delta)
              - N_{\rm leave}(\delta)}{N\,\delta} .
    \label{eq:A-hybrid}
\end{equation}

\begin{figure}[h]
    \centering
    \includegraphics[width=0.95\linewidth]{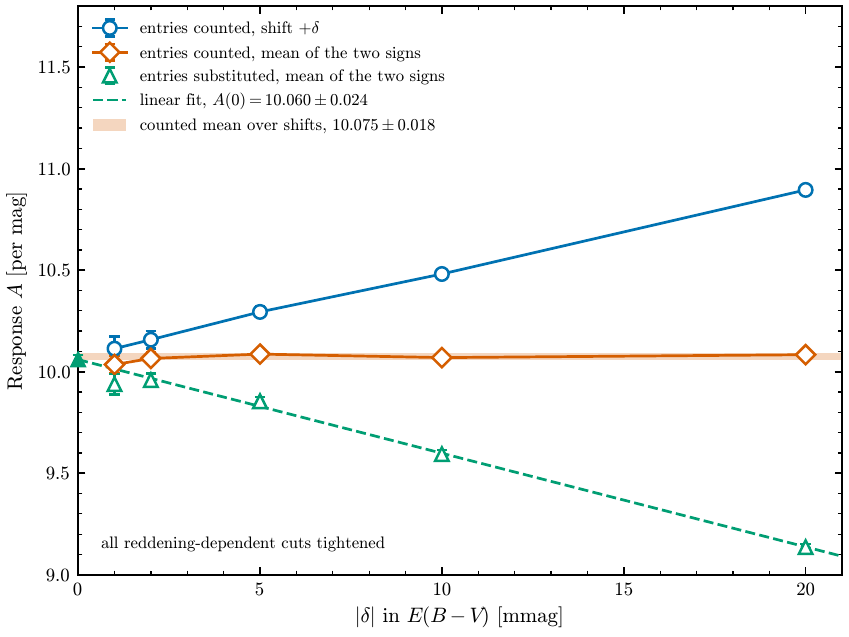}
    \caption{Test of the estimator of Eq.~\eqref{eq:A-hybrid} on a subsample in which every cut that depends on the reddening is tightened, so that all
    entering objects are in the catalog. (Circles) the response with the entering objects counted, for a positive shift. (Diamonds) the same, averaged over the two signs of the shift, and (band) its mean over all
    shifts. (Triangles) the estimator of Eq.~\eqref{eq:A-hybrid}, in which the entering objects are substituted, averaged over the two signs. (Dashed line) linear fit to the triangles, whose value in the small-shift limit
    agrees with the band. The response of this subsample differs from that of the full sample.}
    \label{fig:response-tightened}
\end{figure}

The substitution in the second term is exact only for small shifts. Since the counts rise steeply across the faint limit, the objects just inside it undercount the objects just outside it, by about $0.9\%$ per mmag of shift. Eq.~\eqref{eq:A-hybrid} therefore falls linearly with the size of the shift, from $9.42 \pm 0.04$ per mag at 1 mmag to $8.69 \pm 0.01$ at 20 mmag. We average the two signs and extrapolate linearly to the small-shift limit. To confirm that the fall is a property of the estimator, we repeated the measurement on a subsample in which every cut that depends on the reddening is tightened, so that all entering objects are in the catalog and can be counted (Figure~\ref{fig:response-tightened}). This subsample has its own response, about $10.1$ per mag, which differs from that of the full sample. When the entering objects are counted, the response averaged over the two signs of the shift does not change with the size of the shift. When they are substituted as in Eq.~\eqref{eq:A-hybrid}, the estimate falls linearly with the size of the
shift, and its extrapolation to the small-shift limit, $10.06 \pm 0.02$,
recovers the counted value of $10.08 \pm 0.02$. Errors are from a delete-one jackknife over 100 regions of the footprint.

\subsection{Contribution of each cut}

\begin{table}[htb]
\centering
\small
\begin{tabular}{l c c}
\hline
Cut & Contribution [per mag] & Share of $A$ \\
\hline
$g_{\rm fib} < 24.1$ & $+7.45 \pm 0.01$ & $82.2\%$ \\
$g - r < -1.2\,(r-z) + 1.6$ & $+0.99 \pm 0.01$ & $10.9\%$ \\
$g - r < 0.5\,(r-z) + 0.1$ & $+0.94 \pm 0.01$ & $10.4\%$ \\
$-0.4 < g - r < 0.5$ & $+0.37 \pm 0.01$ & $4.1\%$ \\
$r - z < 1.25$ & $+0.04 \pm 0.01$ & $0.5\%$ \\
$r - z > 0.15$ & $-0.04 \pm 0.01$ & $-0.4\%$ \\
$g_{\rm fib} > 22.6$ & $-0.13 \pm 0.01$ & $-1.5\%$ \\
all other parent cuts & $0$ & $0$ \\
Random Forest classifier & $-0.56 \pm 0.01$ & $-6.2\%$ \\
\hline
Total & $9.07 \pm 0.02$ & $100\%$ \\
\hline
\end{tabular}
\caption{Contribution of each cut to the response of the photometric sample, for a shift of $+10$ mmag. An object that fails more than one cut is assigned to the cut it crosses first as the shift is turned on. The total is lower than the adopted value of Eq.~\eqref{eq:A-value} because of the substitution bias at this shift.}
\label{tab:response-cuts}
\end{table}

Table~\ref{tab:response-cuts} gives the contribution of each cut at a shift of 10 mmag. The faint limit on the $g$-band fiber magnitude carries most of the response. The cuts with a negative contribution are those that objects cross out of the sample when their magnitudes become brighter.

\section{Derivation for a general template}
\label{app:derivation}

This appendix derives the results of Sections~\ref{sec:correction} and
\ref{sec:leftover}. The starting point is the observed overdensity,
Eq.~\eqref{eq:obs}, and the linear correction, Eq.~\eqref{eq:corr-linear}. We
write a general template as
\begin{equation}
    T = G_T + I_T + n_T ,
    \label{eq:template}
\end{equation}
where $G_T$ is its Galactic content, $I_T$ its extragalactic content and $n_T$
its noise, which is independent of everything else. The pair $(G_T, I_T)$
describes the template, while $(\epsilon, I)$ describes the error of the map
used for selection. They coincide only for particular templates, as the
following examples show. A prime denotes the error of a map other than the one
used for selection, and $n$ is the error of the stellar map.

\begin{center}
\begin{tabular}{llll}
\hline
Template & $G_T$ & $I_T$ & $n_T$ \\
\hline
SFD98 $-$ Zhou (selection map $-$ stellar map) & $\epsilon$ & $I$ & $-n$ \\
CSFD $-$ Zhou (CIB-cleaned map $-$ stellar map) & $\epsilon$ & $0$ & $-n$ \\
SFD98 (the selection map itself) & $E_{\rm true} + \epsilon$ & $I$ & $0$ \\
Zhou (a stellar map) & $E_{\rm true}$ & $0$ & $n$ \\
other emission map $-$ Zhou & $\epsilon'$ & $I'$ & $-n$ \\
\hline
\end{tabular}
\end{center}

\subsection{General result}

The slope is fitted to the data,
\begin{equation}
    \beta = \frac{{\rm Cov}\!\left(\delta_g^{\rm obs}, T\right)}{{\rm Var}(T)}
    = \frac{A \left[ {\rm Cov}(\epsilon, G_T) + {\rm Cov}(I, I_T) \right]
            + {\rm Cov}(\delta_g, I_T)}
           {{\rm Var}(G_T) + {\rm Var}(I_T) + {\rm Var}(n_T)} .
    \label{eq:beta-general}
\end{equation}
The first term in the numerator is the response of the selection to the part of
the map error that the template traces. The second, ${\rm Cov}(\delta_g, I_T)$,
is not a selection effect. It is the intrinsic correlation between the galaxies
and the extragalactic signal in the template. The fit cannot tell the two
apart. Inserting Eq.~\eqref{eq:template} in Eq.~\eqref{eq:corr-linear}, the
corrected spectra are
\begin{align}
    C_\ell^{gg,{\rm corr}} &= C_\ell^{gg}
        + 2A\,C_\ell^{gI} - 2\beta\,C_\ell^{g I_T} \nonumber \\
    &\quad + A^2 \left( C_\ell^{\epsilon\epsilon} + C_\ell^{II} \right)
        - 2A\beta \left( C_\ell^{\epsilon G_T} + C_\ell^{I I_T} \right)
        \nonumber \\
    &\quad + \beta^2 \left( C_\ell^{G_T G_T} + C_\ell^{I_T I_T} + N_\ell^{T} \right) ,
        \label{eq:cgg-corr-general} \\
    C_\ell^{\kappa g,{\rm corr}} &= C_\ell^{\kappa g}
        + A\,C_\ell^{\kappa I} - \beta\,C_\ell^{\kappa I_T} ,
        \label{eq:ckg-corr-general}
\end{align}
with $N_\ell^{T}$ the noise power of the template. The cases used in this paper
follow from the content of the template.

\subsection{Reddening-difference template}

The template of Section~\ref{sec:leftover} is $T = \epsilon + I - n$, i.e.,
$G_T = \epsilon$, $I_T = I$ and $n_T = -n$. Eq.~\eqref{eq:beta-general} becomes
\begin{equation}
    \beta
    = A\, \underbrace{\frac{{\rm Var}(\epsilon) + {\rm Var}(I)}{{\rm Var}(T)}}_{\lambda}
      + \underbrace{\frac{{\rm Cov}(\delta_g, I)}{{\rm Var}(T)}}_{\gamma} ,
    \label{eq:app-beta-diff}
\end{equation}
which is Eq.~\eqref{eq:beta-diff}. Here $\lambda$ is the fraction of the
template variance that is map error and not noise, and $\gamma$ is the
regression slope of the true overdensity on the template. The corrected field
is
\begin{equation}
    \delta_g^{\rm corr}
    = \delta_g + (A - \beta)(\epsilon + I) + \beta\,n ,
    \qquad
    A - \beta = A(1-\lambda) - \gamma ,
    \label{eq:app-field-diff}
\end{equation}
and the corrected spectra are
\begin{align}
    C_\ell^{\kappa g,{\rm corr}}
    &= C_\ell^{\kappa g} + (A - \beta)\,C_\ell^{\kappa I} ,
    \label{eq:app-ckg-diff} \\
    C_\ell^{gg,{\rm corr}} &= C_\ell^{gg}
        + 2(A-\beta)\, C_\ell^{gI}
        + (A-\beta)^2 \left( C_\ell^{\epsilon\epsilon} + C_\ell^{II} \right)
        + \beta^2 N_\ell ,
        \label{eq:app-cgg-diff}
\end{align}
with $N_\ell$ the power of the error of the stellar map.
Eq.~\eqref{eq:app-ckg-diff} is the leftover in the lensing cross-spectrum
quoted in Section~\ref{sec:leftover}.

The limit of signal removal is instructive. If the template were noise-free and
purely extragalactic, $T = I$, a fit performed mode by mode would give
$\beta_\ell = A + C_\ell^{gI}/C_\ell^{II}$ and
\begin{equation}
    C_\ell^{gg,{\rm corr}} = C_\ell^{gg}\left( 1 - r_\ell^2 \right) ,
    \qquad
    r_\ell \equiv \frac{C_\ell^{gI}}{\sqrt{C_\ell^{gg}\,C_\ell^{II}}} .
    \label{eq:r2}
\end{equation}
The imprint is removed exactly, together with the fraction $r_\ell^2$ of the
true power that the template can predict. For a difference template this loss
is diluted by the Galactic error and the noise, which dominate ${\rm Var}(T)$.

\subsection{Correlated errors of the two maps}

Eq.~\eqref{eq:app-beta-diff} assumes that the error of the stellar map is
independent of the error of the selection map. Suppose instead that a part of
it follows the Galactic error, $n = \rho\,\epsilon + n_\perp$, with $n_\perp$
independent of $\epsilon$. The shared part cancels in the difference, and the
template becomes $T = (1-\rho)\,\epsilon + I - n_\perp$. For a selection map
without extragalactic signal, the slope and the corrected field are
\begin{equation}
    \beta = \frac{A\,(1-\rho)\,{\rm Var}(\epsilon)}
                 {(1-\rho)^2\,{\rm Var}(\epsilon) + {\rm Var}(n_\perp)} ,
    \qquad
    \delta_g^{\rm corr} = \delta_g
        + \left[ A - \beta\,(1-\rho) \right] \epsilon + \beta\, n_\perp .
    \label{eq:app-rho}
\end{equation}
The fraction of the Galactic imprint that is left is
${\rm Var}(n_\perp) / [(1-\rho)^2\,{\rm Var}(\epsilon) + {\rm Var}(n_\perp)]$.
The shared error lowers the Galactic content of the template by $(1-\rho)^2$
while its noise is unchanged, so more of the imprint is left than for
$\rho = 0$. In the limit $\rho \to 1$ the template contains none of the
Galactic error and the correction removes none of the imprint.

\subsection{The two cases of the forecast}

The forecast of Section~\ref{sec:forecast} uses a correction template without
extragalactic signal, $G_T = \epsilon$ and $I_T = 0$. If the selection map
contains the extragalactic signal $I$, the slope is $\beta = A\,
{\rm Var}(\epsilon)/{\rm Var}(T)$ and
\begin{align}
    C_\ell^{\kappa g,{\rm corr}} &= C_\ell^{\kappa g} + A\,C_\ell^{\kappa I} ,
        \nonumber \\
    C_\ell^{gg,{\rm corr}} &= C_\ell^{gg}
        + 2A\,C_\ell^{gI} + A^2\,C_\ell^{II}
        + (A-\beta)^2\,C_\ell^{\epsilon\epsilon} + \beta^2 N_\ell .
        \label{eq:app-cases}
\end{align}
The terms in $I$ are untouched by the correction. These are the CIB terms of
Eqs.~\eqref{eq:dCkg-cib} and \eqref{eq:dCgg-cib}, and the last two terms are
the power of the residual field of Eq.~\eqref{eq:residual-field}. If the
selection map has no extragalactic signal, the terms in $I$ are absent. The
lensing cross-spectrum is then unchanged, and only the last two terms remain in
the auto-spectrum.

\subsection{Absolute reddening maps}

A regression on the selection map itself has $T = E_{\rm true} + \epsilon + I$,
i.e., $G_T = E_{\rm true} + \epsilon$, $I_T = I$ and $n_T = 0$.
Eq.~\eqref{eq:beta-general} gives
\begin{equation}
    \beta = \frac{A \left[ {\rm Cov}(\epsilon, E_{\rm true})
            + {\rm Var}(\epsilon) + {\rm Var}(I) \right]
            + {\rm Cov}(\delta_g, I)}
           {{\rm Var}(E_{\rm true} + \epsilon) + {\rm Var}(I)} ,
    \label{eq:beta-abs}
\end{equation}
and the corrected field is
\begin{equation}
    \delta_g^{\rm corr} = \delta_g + (A - \beta)(\epsilon + I)
        - \beta\, E_{\rm true} .
    \label{eq:field-abs}
\end{equation}
Two limits show what such a correction can do. If the map error is a pure error
of scale, $\epsilon = k\,E_{\rm true}$ and $I = 0$, then $\beta = A k/(1+k)$ and
$\delta_g^{\rm corr} = \delta_g$, so the imprint is removed exactly. If instead
the error is unrelated to the dust column, ${\rm Cov}(\epsilon, E_{\rm true}) =
0$, the numerator of Eq.~\eqref{eq:beta-abs} contains only the variance of the
error, while the denominator is dominated by the variance of the column. Then
$\beta \ll A$ and almost all of the imprint remains, including the part from
the extragalactic signal. An absolute map is therefore the right template only
for the part of the error that follows the column. The same holds for a
regression on a stellar map, $T = E_{\rm true} + n$, for which
$\beta = A\,{\rm Cov}(\epsilon, E_{\rm true}) / {\rm Var}(T)$.

\subsection{Several templates and nonlinear fits}

With several templates $T_j = G_j + I_j + n_j$ the correction is
$\delta_g^{\rm corr} = \delta_g^{\rm obs} - \sum_j \beta_j T_j$, and the slopes
solve
\begin{equation}
    \sum_k {\rm Cov}(T_j, T_k)\, \beta_k
    = A\, {\rm Cov}(\epsilon + I,\, T_j) + {\rm Cov}(\delta_g, I_j) .
    \label{eq:beta-multi}
\end{equation}
The right-hand side has the same two parts as Eq.~\eqref{eq:beta-general}. A
nonlinear regression, such as a neural network, replaces the linear combination
by a general function of the templates. It can remove more of the imprint than
a linear fit, and by the same mechanism more of the real signal. Two properties
of the linear case carry over. No regression can remove a part of the map error
that is independent of all the templates. Real signal is also removed only
through templates that have extragalactic content, $I_j \neq 0$.

\subsection{Terms beyond first order}
\label{app:second-order}

The correction is a division, so the exact corrected field is
$1 + \delta_g^{\rm corr} = (1 + \delta_g)(1 + A s)/(1 + \beta T)$. With
$A\,r_X = A s - \beta T$ the first-order term of
Eq.~\eqref{eq:residual-field}, the expansion to second order is
\begin{equation}
    \delta_g^{\rm corr} = \delta_g + A\,r_X
        + A\,r_X\,\delta_g - \beta\,T\,A\,r_X .
    \label{eq:second-order}
\end{equation}
The third term couples the residual field to the galaxy field, i.e., the
selection modulates the amplitude of the true clustering. The fourth is
quadratic in the map error. In the two-point function these terms first appear
at third order in the fields. Since the Galactic error and the noise are
independent of $\delta_g$ and have zero mean, the third-order terms that
involve $\delta_g$ come only from the extragalactic signal, through the
bispectra of two galaxy fields with $I$ and of one galaxy field with two
factors of $I$. They vanish for a map without extragalactic signal and for
Gaussian fields. At fourth order the modulation term changes $C_\ell^{gg}$ by a
fraction of order $A^2\,{\rm Var}(r_X)$. For the ELG sample without any
correction this is below $1\%$, with $A = 9.48$ per mag and a map error of at
most $10$~mmag rms. It is small compared to the first-order leftover of
$4$--$13\%$ of the auto-spectrum
(Section~\ref{sec:regression-abs-ebv-residual}).

\section{Details of the leftover measurements}
\label{app:leftover-details}

\subsection{Estimator of the leftover power}
\label{app:PX-estimator}

The leftover power of Eq.~\eqref{eq:PX} is measured on the part of the Legacy
Surveys DR9 footprint that is covered by the Zhou and the Mudur maps, which has
a sky fraction of $0.236$, at a HEALPix resolution of $N_{\rm side} = 256$. We
measure the spectra in seven bands with edges at $\ell = 30$, $50$, $70$,
$100$, $140$, $200$, $280$ and $400$.

The noise power $N_\ell$ of the template is a white-noise level computed from
the statistical errors of the public $g-r$ version of the Zhou map, averaged
over the mask and passed through the same band-power estimator as the spectra.
It is $1\%$ of the power of the template in the lowest band and rises to $33\%$
in the highest. The square of a noisy cross-spectrum is biased high by its
variance, so we subtract the jackknife variance of
$C_\ell^{g_X T_{\rm meas}}$ from the numerator of Eq.~\eqref{eq:PX}. The
fraction $P_X / C_\ell^{g_X g_X}$ uses the auto-spectrum of each run with its
shot noise subtracted. The values quoted for a range of multipoles are weighted
means of the fractions in the bands, with weights given by the squared
signal-to-noise ratio of the auto-spectrum in each band. With these weights,
$90\%$ of the weight of the range $50 \le \ell < 400$ lies at $\ell \ge 100$.
Errors are from a delete-one jackknife over 100 regions.

The second estimate uses the template $T' = E_{\rm CSFD} - E_{\rm Mudur}$ in
the numerator and the cross-spectrum of the two templates in the denominator,
\begin{equation}
    P_X'(\ell) = \frac{\left( C_\ell^{g_X T'} \right)^2}
                      {C_\ell^{T_{\rm meas} T'}} ,
    \label{eq:PX-twomap}
\end{equation}
with the same subtraction of the jackknife variance in the numerator. The noise
of the two stellar maps is independent, so it does not enter the denominator
and no noise model is needed. The galaxy overdensity is correlated with $T'$
only, so this estimate is biased low for the run that used the Mudur map as its
correction template and not for the run that used the Zhou map. In all
measurement templates we use the version of the Mudur map that is rescaled in
declination (Section~\ref{sec:dust}).

\subsection{Shared calibration of the two stellar maps}
\label{app:shared-calibration}

The two stellar maps are built from different stars and with different methods,
but they are not independent. The Zhou map uses the colors of stars in the
Legacy Surveys photometry, whose zero points are tied to Pan-STARRS1
\cite{rongpu-desi-dustmap-zhou}. The Mudur map is built from Pan-STARRS1 and
2MASS photometry \cite{nayantara-mudur-dust}. An error in the Pan-STARRS1
calibration can therefore enter both maps, and it also enters the photometry on
which the galaxies are selected.

\subsection{Tests with the H\,\textsc{I} map and with the two versions of the Zhou map}
\label{app:calibration-tests}

\paragraph{Comparison with the H\,\textsc{I} map.}
We compare the three templates $E_{\rm CSFD} - E_{\rm Zhou}$,
$E_{\rm CSFD} - E_{\rm Mudur}$ and $E_{\rm CSFD} - E_{\rm HI}$ on the part of
the footprint where the H\,\textsc{i} map is valid, which has a sky fraction of
$0.16$. If the three reference maps had independent errors, the covariance of
any two templates would be the variance of the error of CSFD, and the
covariances would be equal. We find that the covariance of the two stellar
templates is $25.3 \pm 5.5$~mmag$^2$, while the covariance of the Zhou template
with the H\,\textsc{i} template is $8.1 \pm 2.7$~mmag$^2$. The difference is
$17.3 \pm 6.3$~mmag$^2$, which is $17\%$ of the variance of
$E_{\rm CSFD} - E_{\rm Zhou}$. The two stellar maps therefore agree with each
other more closely than either agrees with the H\,\textsc{i} map, at a
significance of $2.7\sigma$. This can mean that the two stellar maps share an
error, or that the H\,\textsc{i} map shares an error with SFD98. Our data do
not distinguish the two.

\begin{figure}[h]
    \centering
    \includegraphics[width=\linewidth]{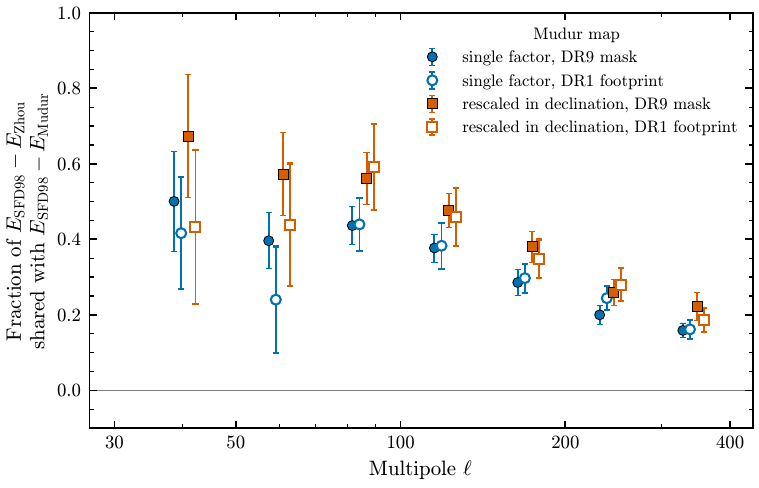}
\caption{Fraction of the power of the template $E_{\rm SFD98} - E_{\rm Zhou}$ that it shares with $E_{\rm SFD98} - E_{\rm Mudur}$, i.e., the cross-spectrum of the two templates divided by the auto-spectrum of the first. Circles use the Mudur map scaled to CSFD by a single factor, and squares use the version rescaled in declination (Section~\ref{sec:dust}). Filled symbols are for the part of the DR9 mask covered by the Zhou and Mudur maps, on which we measure the leftover power (Appendix~\ref{app:PX-estimator}), and open symbols are for the DR1 footprint, on which we test the DESI weights (Section~\ref{sec:template-limits}). The noise of the Zhou map is not subtracted from the auto-spectrum. Errors are from a delete-one jackknife over 100 regions.}
    \label{fig:lambda-scale}
\end{figure}

\paragraph{Shared fraction of the difference templates.}
Figure~\ref{fig:lambda-scale} shows how much the reddening-difference templates built with the two stellar maps have in common, as a function of scale. For the version of the Mudur map rescaled in declination, on the DR1 footprint, the shared fraction is about a half at the largest scales and falls below $0.2$ at the smallest, and it is about a quarter over all scales. The fall toward small scales is expected from the noise of the Zhou map, which is a growing share of the power of its template at high multipoles. With the Mudur map scaled by a single factor, the shared fraction is lower in every band on both footprints, which is the comparison used in Section~\ref{sec:dust} to place the drift of the reddening scale in the Mudur map.

\paragraph{The two versions of the Zhou map.}
The direct test of Section~\ref{sec:template-limits} uses the difference
between the $g-r$ and the $r-z$ versions of the Zhou map. We fit the overdensity
of the DR9 photometric sample without imaging weights against this difference
over $30 \le \ell < 400$, after subtracting the noise of both versions from the
power of the difference. The slope is $-2.75 \pm 0.30$ per mag. For comparison
we predict the slope for independent zero-point errors in $g$, $r$ and $z$ with
rms values of $4.7$, $3.7$ and $4.4$~mmag
\cite{rongpu-desi-dustmap-zhou}, combined with the measured
response of the sample to each band (Table~\ref{tab:response-summary}). This
gives $-2.41$ per mag, consistent with the measurement. On the part of the
footprint where the H\,\textsc{i} map is valid, i.e., the sky with the lowest
column of gas and dust, the slope is $-2.77 \pm 0.42$ per mag. The signal is
therefore also present where a variation of the extinction curve would have
little effect.

\section{Modeling of the angular power spectra}
\label{app:theory}

The setup of this paper follows a photometric survey-style analysis where the
projected power spectra are modeled. In the following sections we discuss how
the 3D power spectrum is mapped into the projected space, and how the projected
spectra are linked to the different cosmological parameters of interest.

\subsection{Galaxy and CMB-lensing power spectra}
\label{sec:theory-general}

The angular power spectrum of two fields $A$ and $B$ is
\begin{equation}
    C_\ell^{AB} = \frac{2}{\pi} \int k^2\, dk\; P_{\rm lin}(k)\,
    \Delta_\ell^{A}(k)\, \Delta_\ell^{B}(k) ,
    \label{eq:cl-exact}
\end{equation}
with $P_{\rm lin}(k)$ the linear matter power spectrum today. For clarity we
write the transfer functions with a scale-independent growth factor $D(z)$; the
numerical calculation uses the full transfer functions. For the galaxies of
tomographic bin $i$, the transfer function has a term from the intrinsic
clustering and a term from redshift-space distortions (RSD),
\begin{equation}
    \Delta_\ell^{g_i}(k) = \int d\chi\, W_{g_i}(\chi)\, D(z)
    \left[ b_i(k, z)\, j_\ell(k\chi) - f(z)\, j_\ell''(k\chi) \right] ,
    \label{eq:transfer-g}
\end{equation}
where $W_{g_i}(\chi) = n_i(z)\, dz/d\chi$ with $n_i(z)$ the normalized redshift
distribution, $b_i$ is the galaxy bias, $f$ is the linear growth rate, and
$j_\ell$ is a spherical Bessel function. For CMB lensing,
\begin{align}
    \Delta_\ell^{\kappa}(k) &= \frac{\ell(\ell+1)}{k^2}
    \int d\chi\, \frac{W_\kappa(\chi)}{\chi^2}\, D(z)\, j_\ell(k\chi) ,
    \label{eq:transfer-k} \\
    W_\kappa(\chi) &= \frac{3\,\Omega_m H_0^2}{2 c^2}\,
    \frac{\chi}{a(\chi)}\, \frac{\chi_* - \chi}{\chi_*} ,
    \label{eq:Wkappa}
\end{align}
with $\chi_*$ the distance to last scattering. We do not include lensing
magnification in the model. As a check we added it to the forecast for
$f_{\rm NL}$ with a count slope of $s = 0.359$ \cite{lsst-desc-srd}, where it
changes the bias by at most $0.03\sigma$.

When the kernels are broad compared with the scales of interest, the Limber
approximation replaces Eq.~\eqref{eq:cl-exact} by
\begin{equation}
    C_\ell^{AB} \simeq \int \frac{d\chi}{\chi^2}\,
    W_A(\chi)\, W_B(\chi)\,
    P_{AB}\!\left(k = \frac{\ell + 1/2}{\chi},\, z\right) ,
    \label{eq:limber}
\end{equation}
where $P_{AB}$ is the three-dimensional cross-spectrum of the two tracers,
including their biases. In this limit the RSD term is suppressed. For the
auto-spectra and the cross-spectra with lensing of the bins used here,
Eq.~\eqref{eq:limber} agrees with the exact calculation to better than $1\%$ at
$\ell > 100$ and to about $2\%$ at $30 \le \ell \le 100$. The response of the
spectra to $f_{\rm NL}$ is less well reproduced, and differs by up to $13\%$ at
$\ell = 30$.

\subsection{Amplitude and growth: $\sigma_8$, $\Omega_m$ and $\Sigma m_\nu$}
\label{sec:theory-amplitude}

On linear scales, and in the Limber limit, the two spectra of bin $i$ scale as
\begin{equation}
    C_\ell^{g_i g_i} \propto b_i^2\, \sigma_8^2\, D^2(z_i) ,
    \qquad
    C_\ell^{\kappa g_i} \propto b_i\, \Omega_m\, \sigma_8^2\, D^2(z_i) ,
    \label{eq:amp-scaling}
\end{equation}
so that the combination
\begin{equation}
    \frac{C_\ell^{\kappa g_i}}{\sqrt{C_\ell^{g_i g_i}}}
    \propto \Omega_m\, \sigma_8\, D(z_i)
    \label{eq:amp-ratio}
\end{equation}
does not depend on the galaxy bias. An excess in $C_\ell^{g_i g_i}$ alone that
has the same shape as the spectrum is absorbed as a larger bias and lowers the
inferred amplitude, while an excess in $C_\ell^{\kappa g_i}$ raises it. An
excess with a different shape, such as the leftover power of the imprint, also
changes the parameters that set the shape of the spectra
(Section~\ref{sec:results-amplitude}). The matter density enters through this
shape and through the distances in the kernels. Massive neutrinos suppress the
growth of structure below their free-streaming scale, which lowers the
amplitude measured at low redshift relative to the primordial amplitude fixed
by the CMB. With several redshift bins, Eq.~\eqref{eq:amp-ratio} traces the
growth history, which is what constrains $\Sigma m_\nu$.

For these parameters we use the Limber form, Eq.~\eqref{eq:limber}, at
$\ell \ge 30$, without the RSD term, with the nonlinear matter power spectrum of \textsc{halofit} \cite{takahashi-halofit-2012} as implemented in \textsc{camb}. The neutrino mass is carried by one massive eigenstate. The
galaxies trace the field of cold dark matter and baryons, while the lensing
kernel uses the total matter.

\subsection{Primordial non-Gaussianity: $f_{\rm NL}$}
\label{sec:theory-fnl}

Local primordial non-Gaussianity makes the galaxy bias scale dependent,
\begin{equation}
    b_i(k, z) = b_i + f_{\rm NL}\, (b_i - p)\,
    \frac{3\, \delta_{\rm crit}\, \Omega_m H_0^2}
         {c^2\, k^2\, T(k)\, D_{\rm md}(z)} ,
    \label{eq:bias-fnl}
\end{equation}
where $\delta_{\rm crit} = 1.686$, $T(k)$ is the matter transfer function
normalized to unity on large scales, and $D_{\rm md}(z)$ is the growth factor
normalized to the scale factor during matter domination. We set $p = 0.55$, and
not the usual $p = 1$. We choose this value based on studies that have shown
that galaxies selected by stellar mass prefer it in hydrodynamic simulations
\cite{barreira-universality-p}. While a color selection does not fully map to a
selection by stellar mass, we use this number as a proxy in our paper. The
second term grows as $k^{-2}$, so the signal is concentrated at the lowest
multipoles of $C_\ell^{g_i g_i}$ and $C_\ell^{\kappa g_i}$.

These are the multipoles where the Limber approximation fails and where the RSD
term in Eq.~\eqref{eq:transfer-g} has the same $\ell^{-2}$ dependence as the
signal. For $f_{\rm NL}$ we therefore use the exact expression,
Eq.~\eqref{eq:cl-exact}, at all multipoles, with the scale-dependent bias of
Eq.~\eqref{eq:bias-fnl} inserted in Eq.~\eqref{eq:transfer-g} and with the RSD
term included. The density terms use the nonlinear matter power spectrum, and
the terms proportional to $f_{\rm NL}$ and the RSD terms use the linear one. We
use $C_\ell^{g_i g_i}$ at $\ell \ge 30$ and $C_\ell^{\kappa g_i}$ at
$\ell \ge 8$. We also report the Limber result for comparison
(Table~\ref{tab:fnl-robustness}).

\subsection{Modeling of the galaxy bias}
\label{sec:galaxy-bias}

We use a linear bias with one free amplitude $b_i$ per lens bin. The fiducial
values follow the DESC SRD \cite{lsst-desc-srd}, $b(z) = 1.05/D(z)$ for Year~1
and $0.95/D(z)$ for Year~10, with $D(z)$ normalized to one today, evaluated at
the mean redshift of each bin. As a test of robustness we add one parameter per
bin that makes the bias scale dependent on small scales,
\begin{equation}
    b_i(k) = b_i \left( 1 + b_{k,i}\, k^2 \right) ,
    \label{eq:bias-k2}
\end{equation}
with fiducial $b_{k,i} = 0$. This is an effective term, not a full model of
nonlinear bias. It lets the model change the ratio of small-scale to
large-scale power (Section~\ref{sec:results-fnl}).

\subsection{From theory to measured spectra}
\label{sec:pseudo-cl}

The spectra above are for the full sky. A survey mask couples the multipoles,
so the spectrum measured on the masked sky is related to the true one by the
mode-coupling matrix of the mask \cite{hivon-master}. We compute and invert
this matrix with \textsc{NaMaster} \cite{namaster}, and all measured spectra in
this paper are binned and decoupled band powers.

\subsection{The Galactic term for each correction}
\label{app:amplitude-corrections}

Table~\ref{tab:amplitude-corrections} gives the Galactic term in the shifts of
$\sigma_8$, $\Omega_m$ and $\Sigma m_\nu$ for every correction, to complement
the results for the sample corrected with SFD98 and for the DESI weights in
Section~\ref{sec:results-amplitude}.

\begin{table}[ht]
\centering
\small
\begin{tabular}{lccc}
\hline
Correction & $\Delta\sigma_8/\sigma$ & $\Delta\Omega_m/\sigma$ & $\Delta\Sigma m_\nu/\sigma$ \\
\hline
\multicolumn{4}{l}{\textit{Year 1}} \\
No weights & $+1.19 \pm 0.29$ & $-2.74 \pm 0.58$ & $+1.59 \pm 0.36$ \\
SFD98 & $+0.35 \pm 0.10$ & $-0.97 \pm 0.20$ & $+0.50 \pm 0.12$ \\
CSFD & $+0.33 \pm 0.09$ & $-0.94 \pm 0.20$ & $+0.47 \pm 0.11$ \\
GNILC & $+0.23 \pm 0.08$ & $-0.67 \pm 0.15$ & $+0.34 \pm 0.09$ \\
Mudur & $+0.25 \pm 0.07$ & $-0.74 \pm 0.15$ & $+0.38 \pm 0.08$ \\
Zhou & $+0.20 \pm 0.05$ & $-0.61 \pm 0.11$ & $+0.27 \pm 0.07$ \\
Zhou, two stellar maps & $+0.23 \pm 0.14$ & $-0.79 \pm 0.27$ & $+0.30 \pm 0.18$ \\
DESI weights, $R = 0.33$ & $+0.13 \pm 0.03$ & $-0.30 \pm 0.06$ & $+0.17 \pm 0.04$ \\
DESI weights, $R = 0.50$ & $+0.30 \pm 0.07$ & $-0.68 \pm 0.15$ & $+0.40 \pm 0.09$ \\
\hline
\multicolumn{4}{l}{\textit{Year 10}} \\
No weights & $+1.32 \pm 0.37$ & $-3.67 \pm 0.79$ & $+2.64 \pm 0.60$ \\
SFD98 & $+0.27 \pm 0.13$ & $-1.28 \pm 0.27$ & $+0.81 \pm 0.20$ \\
CSFD & $+0.24 \pm 0.13$ & $-1.23 \pm 0.26$ & $+0.77 \pm 0.19$ \\
GNILC & $+0.17 \pm 0.11$ & $-0.87 \pm 0.20$ & $+0.55 \pm 0.14$ \\
Mudur & $+0.18 \pm 0.11$ & $-0.96 \pm 0.20$ & $+0.62 \pm 0.13$ \\
Zhou & $+0.15 \pm 0.08$ & $-0.80 \pm 0.15$ & $+0.42 \pm 0.11$ \\
Zhou, two stellar maps & $+0.13 \pm 0.22$ & $-1.03 \pm 0.37$ & $+0.45 \pm 0.31$ \\
DESI weights, $R = 0.33$ & $+0.14 \pm 0.04$ & $-0.40 \pm 0.09$ & $+0.29 \pm 0.07$ \\
DESI weights, $R = 0.50$ & $+0.33 \pm 0.09$ & $-0.92 \pm 0.20$ & $+0.66 \pm 0.15$ \\
\hline
\end{tabular}
\caption{The Galactic term in the shifts of $\sigma_8$, $\Omega_m$ and
$\Sigma m_\nu$ for each correction. The prior on the width of the lens redshift
distributions is the SRD requirement for the source sample, $0.006$ in Year~1
and $0.003$ in Year~10. Uncertainties propagate the jackknife uncertainty of
the measured leftover power. They are strongly correlated between rows, since
all entries share the same galaxy map. The row with two stellar maps uses the
second estimate of the leftover power for the sample corrected with the Zhou
map (Section~\ref{sec:regression-abs-ebv-residual}). The CIB terms are the same
for every row and are given in Table~\ref{tab:amplitude-terms}.}
\label{tab:amplitude-corrections}
\end{table}

\clearpage
\bibliographystyle{JHEP}
\bibliography{biblio.bib}

\end{document}